\documentclass{article}
\usepackage[utf8]{inputenc}

\usepackage{url,hyperref,lineno,microtype}
\usepackage[onehalfspacing]{setspace}
\usepackage{amsmath,amssymb,bbm}
\usepackage{multirow,float}
\usepackage{blindtext}
\usepackage{amsthm}
\usepackage{amsfonts}
\usepackage{float}
\usepackage{graphicx}
\usepackage{mathtools}
\usepackage{caption}
\usepackage{subcaption}
\usepackage{enumitem}
\usepackage{bbold}
\usepackage{soul,xcolor}
\usepackage{authblk}
\usepackage{geometry}
\usepackage{hyperref}
\usepackage{caption}
\usepackage[toc,page]{appendix} 
\usepackage[style=numeric-comp,backend=biber,sorting=none]{biblatex}
\newcommand{\1}{\mathbbm{1}}

\newcommand{\cI}{\mathcal{I}}

\begin{document}
	
\begin{flushleft}
	{\Large
		\textbf\newline{A Simple Modeling Framework For Prediction In The Human Glucose-Insulin System} % Please use "sentence case" for title and headings (capitalize only the first word in a title (or heading), the first word in a subtitle (or subheading), and any proper nouns).
	}
	\newline
	\\
	Melike Sirlanci\textsuperscript{5\textcurrency*},
	Matthew E. Levine\textsuperscript{5},
	Cecilia Low Wang\textsuperscript{6}
	David J. Albers\textsuperscript{1,2,3,4},
	Andrew M. Stuart\textsuperscript{5},
	\\
	\bigskip
	\textbf{1} Department of Biomedical Informatics, School of Medicine, University of Colorado Anschutz Medical Campus, Aurora, CO, USA
	\\
	\textbf{2} Department of Bioengineering, College of Engineering, Design, and Computing, Aurora, CO, USA
	\\
	\textbf{3} Department of Biostatistics \& Informatics, Colorado School of Public Health, Aurora, CO, USA
	\\
	\textbf{4} Department of Biomedical Informatics, Columbia University, New York, NY, USA
	\\
	\textbf{5} Department of Computing and Mathematical Sciences, California Institute of Technology, Pasadena, CA, USA
	\\
	\textbf{6} Division of Endocrinology, Metabolism and Diabetes, Department of Medicine, School of Medicine, University of Colorado Anschutz Medical Campus, Aurora, CO, USA
	\\
	\bigskip
	
	% Insert additional author notes using the symbols described below. Insert symbol callouts after author names as necessary.
	%
	% Remove or comment out the author notes below if they aren't used.
	%
	% Primary Equal Contribution Note
	%\Yinyang These authors contributed equally to this work.
	
	% Additional Equal Contribution Note
	% Also use this double-dagger symbol for special authorship notes, such as senior authorship.
	%\ddag These authors also contributed equally to this work.
	
	\textcurrency Current Address: Department of Biomedical Informatics, School of Medicine, University of Colorado Anschutz Medical Campus, Aurora, CO, USA % change symbol to "\textcurrency a" if more than one current address note
	% \textcurrency b Insert second current address
	% \textcurrency c Insert third current address
	
	
	% Group/Consortium Author Note
	%\textpilcrow Membership list can be found in the Acknowledgments section.
	
	% Use the asterisk to denote corresponding authorship and provide email address in note below.
	*melike.sirlanci@cuanschutz.edu
	
\end{flushleft}

\begin{abstract}

%%% Leave the Abstract empty if your article does not require one, please see the Summary Table for full details.
In this paper, we build a new, simple, and interpretable mathematical model to estimate and forecast physiology related to the human glucose-insulin system, constrained by available data. By constructing a simple yet flexible model class with interpretable parameters, this general model can be specialized to work in different settings, such as type 2 diabetes mellitus (T2DM) and intensive care unit (ICU); different choices of appropriate model functions describing uptake of nutrition and removal of glucose differentiate between the models. In addition to data-driven decision-making, the model has the potential also to be useful for the basic quantification of endocrine physiology. In both cases, the available data is sparse and collected in clinical settings, major factors that have constrained our model choice to the simple form adopted.

The model has the form of a linear stochastic differential equation (SDE) to describe the evolution of the BG level. The model includes a term quantifying glucose removal from the bloodstream through the regulation system of the human body and two other terms representing the effect of nutrition and externally delivered insulin. The stochastic fluctuations encapsulate model error necessitated by the simple model form and enable flexible incorporation of data. The parameters entering the equation must be learned in a patient-specific fashion, leading to personalized models. We present experimental results on patient-specific parameter estimation and future BG level forecasting in T2DM and ICU settings. The resulting model leads to the prediction of the BG level as an expected value accompanied by a band around this value which accounts for uncertainties in the prediction. Such predictions, then, have the potential for use as part of control systems that are robust to model imperfections and noisy data. Finally, the model's predictive capability is compared with two different models built explicitly for T2DM and ICU contexts. Our simple model also shows clear advantages over the more complex models in terms of parameter inference and identifiability, as well as controllability, stemming from its linearity.

\end{abstract}

\section{Introduction}

Broadly speaking mathematical models of human physiology may serve
one of two purposes: elucidation of the detailed mechanisms which
comprise the complex systems underlying observed physiology; or
prediction of outcomes from the complex system, for
the purposes of medical intervention to ameliorate undesirable
outcomes. In principle, these two objectives interact: a model
which explains the detailed mechanisms, if physiologically accurate
and compatible with observed data, will of course be good for prediction.
However, human physiological data are often too sparse for use in resolving
high-fidelity physiological details; moreover, this sparsity can induce severe
model unidentifiability that impedes inference efficiency and results in
suboptimal predictive performance. One approach to mitigate unidentifiability issues with high-fidelity models
is to constrain inference. However, in this paper, we focus on how model reduction and stochastic closure techniques
can be applied to physiologic models to make them more identifiable from available data.
This, of course, comes with a cost of reduced fidelity; however, we find that this tradeoff often sides with model simplicity, especially when data are low-fidelity (i.e. sparse and noisy) and the underlying system is not fully understood (i.e. available ``high-fidelity" models have substantial inadequacies).
The human glucose-insulin system provides an important example of this
challenge because in many settings, insulin\footnote[1]{Here, we refer to internal insulin levels such as plasma insulin or interstitial insulin; not the dosages of medication.}---a dominant state variable---is rarely measured.

The objective of the work presented here is to
distill existing mechanistic models of the human endocrine system
into an interpretable model of human glucose dynamics that is identifiable from real-world clinical data.
We do this by approximating the insulin's glycemic regulation as an Ornstein-Uhlenbeck process
(a linear stochastic differential equation with exponential mean-reversion),
then further introducing forcing terms that parameterize exogenous effects of nutrition and medication. The resulting model represents the mean blood glucose (BG) behavior and a confidence region quantifying the amplitude of the BG oscillations. These confidence regions can also be used to quantify the uncertainty in the mean BG behavior. We then evaluate the predictive performance of this simple model on clinical datasets in an outpatient type 2 diabetes setting and
an inpatient intensive care unit setting.
We compare its predictive performance with state-of-the-art predictions given by a physiologically constrained inference machinery paired with popular mechanistic models of the glucose-insulin system (from which our reduced model was inspired).

The key finding from this work is \emph{non-inferior} predictive capacity of our simple linear stochastic model when compared to
higher complexity non-linear models.
This indicates that the severity of our clinical data constraints
prevented us from extracting additional expressivity from the
non-linear models beyond the simple dynamics encoded by a forced linear SDE.
Alternatively, it may be that the additional expressivity of the non-linear models is not of the right type, and thus does not offer much additional predictive advantage (despite having clear mechanistic validity).

Researchers have developed various mathematical models ranging from extremely simple to highly complex, using ordinary differential equations (ODEs) and machine learning (ML) to predict and describe human glucose metabolism. We discuss these efforts organized according to model usage.

Some mechanistic models are developed to investigate a specific phenomenon of the glucose-insulin system such as to understand the different phases of insulin secretion with respect to different glucose stimulation patterns, to estimate insulin sensitivity in the intravenous glucose tolerance test (IVGTT) setting, and to elucidate the cause of the ultradian (long-period) oscillations of insulin and glucose,  \cite{grodsky72,bergman79,li2012range,shi2020analysis,sturis1991computer,li2007analysis,liu2009molecular}. Others have developed models by clinically minded motivations to describe $\beta$-cell mass, glucose, and insulin dynamics and to investigate T2DM pathophysiology, \cite{topp2000model,bertuzzi07,ha2015mathematical,goel2015insulin}. Some researchers developed models to describe the underlying system in more detailed way such as the events that occur during oral glucose ingestion \cite{dalla07,lehmann92}, or relevant organ systems, \cite{eddy2003archimedes}. A nice review of the models developed for clinical and physiological investigation BG homeostasis and T2DM can be found in \cite{mari2020mathematical}.

There are also machine learning models developed to understand model phenotypic and health care process differences and to predict T2DM development, \cite{kahn1994contribution,hripcsak2011exploiting,albers2012using,albers2012estimation,albers2012population,hripcsak2013correlating,albers2014dynamical,lee2015identification,luo2016automatically,beaulieu2017machine,abbas2019predicting,ismail2022type}.

Researchers have developed mechanistic models to address challenges including fast evolution of the underlying system (parameter variation in time), wide variation in clinical response within and between patients, sparse measurements, and concerns about safety issues with the goals of prediction and control of BG levels, \cite{sedigh2012data,vanherpe06,lin2004adaptive,lin2011physiological,pritchard2017modeling,knab2016virtual,roy2010phenomenological,knab2015zone,vilkhovoy2014control,haverbeke2008nonlinear,parker2018impact}. Others developed stochastic (mechanistic) models with the same purpose, \cite{zhang2016data,davidson2019multi,lin2008stochastic,lin2006stochastic,le2009blood,davidson2020virtual,duun2013model}.

Glucose control based on mechanistic modeling is the focus of the artificial pancreas project in the type 1 diabetes mellitus (T1DM) setting and many models are developed for this purpose, \cite{brunetti2003artificial,parker2001intravenous,cobelli1983evaluation,fabietti2007clinical,fabietti2006control,kovatchev2009silico,parker2000robust}. A comprehensive range of BG control algorithms can be found in \cite{chee2007closed}. Finally, other researchers conducted clinical trials to compare the efficacy between different closed-loop artificial pancreas systems and sensor-assisted pump therapy for T1DM patients, \cite{benhamou2019closed,brown2019six,bruttomesso2019toward,tauschmann2018closed,thabit2015home}.

ML approaches have been proposed in pure prediction tasks such as predicting next glucose values or hypoglycemia. For these purposes, some researchers used classification methods and neural network models,  \cite{sudharsan2014hypoglycemia,murata2004probabilistic,gibson2013development,zeevi2015personalized,beverlin2011algorithm,rollins2010free,zitar2005towards}, while others used ARIMA (auto-regressive integrated moving average) and linear regression models, \cite{martinovic2019modelling,yang2018arima,montaser2017stochastic,reifman2007predictive,bremer1999blood,gani2009predicting,sparacino2007glucose,zhang15}.

Finally, in \cite{miller2020learning}, the authors developed a hybrid model balancing a physiological and statistical model of glucose-insulin dynamics to forecast long-term BG levels of T1DM patients based on real-world data, showing  the possibility of outperforming the forecasting of BG levels obtained by either pure physiological or pure statistical models alone.

Patient-centered disease self-management is a crucial tool to improve health condition of patients focusing on their needs, life style, and preferences. Some researchers investigated techniques for effective self glycemic management and developed computational model-based decision support tools for T2DM patients, \cite{cavanaugh2008association,gibson2012efficacy,mamykina2016data,mamykina2017personal,desai2019personal,desai2018pictures,albers2018mechanistic,lum2019decision,glachs2021predictive,mitchell2021enabling}.

In all of the models discussed above, parameter estimation plays a vital role in the accuracy of predictions. Parameters are rarely directly measurable, and their values will vary from one patient to another. There are two overarching approaches to estimating parameters, optimization where a model-data mismatch is minimized to determine parameters \cite{engl1996regularization}, and the Bayesian approach \cite{kaipio2006statistical} where the distribution of the parameters, given the data and given the assumed (noisy) model-data framework, is computed. Researchers used various approaches for parameter estimation. The most common approaches are the standard least squares optimization, \cite{vanherpe06,wu16}, nonlinear least squares optimization, \cite{vanherpe07}, and Bayesian approach to estimate both time-invariant and time-varying model parameters, \cite{hovorka08}.

\textit{Our contribution} in this paper is summarized below.

\begin{itemize}
	
	\item We describe a simple, interpretable,
	modeling framework limited to states and parameters that are directly observable or inferable from data for prediction
	within the human glucose-insulin system, based on a continuous time linear, Gaussian, stochastic differential equation (SDE) for glucose dynamics,
	in which the effect of insulin appears parametrically.
	
	\item We completely describe the inference machinery necessary---in a data assimilation and inverse problems framework---to estimate a SDE model of glucose dynamics with real-world data.
	
	\item The framework is sufficiently general to be usable within the ICU, T2DM, and potentially T1DM settings.
	
	\item The solution of the model can be obtained analytically, which means that it does not require numerical solver and the prediction could quickly be obtained in an online setting. Hence the model could easily be used in any platform for prediction based on real-world data.
	
	\item We demonstrate, in a train-test set-up, that the models are able
	to fit individual patients with reasonable accuracy; both ICU and T2DM data are used. The test framework we use is a predictive one laying the foundations for future control methodologies.
	
	\item Comparison of the data fitting for T2DM and ICU patients reveals interesting structural differences in their glucose regulation.
	
	\item We make a comparison of the predictive power of our stochastic modeling
	framework with that of more sophisticated models developed for both
	T2DM and the ICU, demonstrating that the simple stochastic approach is at least as accurate as these models in both settings.
	
\end{itemize}

In Section \ref{model_construct}, we introduce the general continuous-time mathematical model that describes the human glucose regulatory system. Then, in Section \ref{event_time_model},
we introduce the specific versions of this model relevant in T2DM and ICU settings. The two model classes all derive from a single general model, and
differ according to how nutrition uptake and glucose removal are represented.
In Section \ref{param_est}, we construct the framework for stating the parameter estimation problem and its solution. In Section \ref{exp_design}, we describe the datasets, the experiments we design for parameter estimation and forecasting, and the methods we use for parameter estimation and forecasting for the T2DM and ICU settings. Section \ref{num_results} presents the experimental results on parameter estimation and forecasting along with some uncertainty quantification (UQ) results separately for T2DM and ICU settings. Finally, in Section \ref{conc}, we make some concluding remarks and discuss future directions that we intend to pursue.

\section{Materials and Methods}

In this section, we describe the clinical settings of interest, our model construction, the mathematical techniques, and the experimental design.

\subsection{Clinical Settings of Interest}\label{clinical_setting}

The model that we develop can be used in T2DM and ICU
settings with appropriate adjustments, which will be presented
in Section \ref{model_construct}. In this section, we describe
the respective clinical settings that structure our modeling approach.

\subsubsection{Type 2 Diabetes Mellitus (T2DM)}

Glucose dynamics in this setting are driven by a combination of diet,
activity, medication, and internal physiology. Here, we specifically focus
on modeling the effect of carbohydrate intake on glycemic levels of
people with T2DM. The self-monitoring T2DM dataset is from a previous
prospective self-management trial. It contains the carbohydrate
intake in the meals and 1-2 BG measurements collected before and
after the meals with the corresponding timing of each event. None
of the T2DM patients in our dataset took exogenous insulin to
control their BG level. This means that the carbohydrate intake is
the only input to our model.

The BG levels of T2DM patients show non-stationary behavior over
long time-scales reflecting gradual changes in the health condition.
The observable BG behavior change could occur over time-scales
on the order of months. Therefore, it is possible to capture system
dynamics with a mechanistic model over shorter time intervals, i.e.,
weeks, and use that information to forecast BG levels over the
following few weeks. This type of predictive tool would be beneficial
for T2DM patients in managing their disease. Thus, in this setting,
we design the predictive framework to provide decision aid to T2DM
patients in self-management.

\subsubsection{Intensive Care Unit (ICU)}

In the ICU setting, glucose dynamics are given by a combination of
changing patient physiologic state, nutrition (delivered intravenously
and enterally through a feeding tube that runs to the gut), and insulin
delivery. In ICU setting, on average, 8-10\% of the ICU patients are
diabetic and only 5\% of those are T1DM patients. However, more than
90\% of ICU patients require glycemic management and 10-20\% of
them experience a hypoglycemic event over the course of management.
Consequently, regardless of being diabetic or non-diabetic, they are
typically given IV insulin to control BG levels.

Patients in the ICU typically have much more volatile physiological
dynamics for at least three reasons: glycemic dynamics under continuous
feeding are oscillatory, the patients are acutely ill and their health state
changes quickly because of their disease state, and the patients are
constantly being intervened on to help them heal. Therefore they exhibit
BG time series that are often non-stationary in
complex ways and on different time scales. On slower time scales,
patients eventually leave the ICU because their health either improves
or declines. But there can be fast time scale changes too due to
interventions and/or sudden health-related events, such as a stroke.
These health changes will lead to changes in the best-fit parameters
of the model; in other words the patient-specific model itself may
change abruptly, in contrast to the T2DM case.

The retrospective ICU dataset is extracted from the Columbia University
Medical Center Clinical Data Warehouse. It contains carbohydrate rate
through the enteral feeding tube, IV insulin rate, BG measurements, and
the timing of all these events. It is important to emphasize that we do not
have plasma insulin or interstitial insulin rate, as they are collected rarely.
The carbohydrate and IV insulin rates are the
inputs to the model. Considering the highly non-stationary behavior of the
system, the BG measurements are sparse, at most 15 measurements per day.
In this case, we aim this predictive framework to be used as a clinical
decision support tool in the ICU setting.

\subsection{Model construction}\label{model_construct}
To begin construction of a simple, one-state model for glucose dynamics, we
first consider the classical two-state Bergman \cite{bergman79} equations:
\begin{subequations}
	\label{berg}
	\begin{align}
		\dot{G} &= m_\textrm{external}(t) + f_\textrm{HGP}(G) - (c + s_I I) G \\
		\dot{I} &=
		I_\textrm{external}(t) + \beta f_\textrm{ISR}(G) - kI.
	\end{align}
\end{subequations}
Here, $G$ denotes plasma glucose concentration and $I$ denotes plasma insulin concentration.
External inputs of nutrition and insulin are given by $m_\textrm{external}(t), I_\textrm{external}(t)$, respectively.
The insulin dynamics, beyond external forcing, are primarily governed by a glucose-dependent secretion rate $f_\textrm{ISR}(G)$,
insulin-producing beta-cell mass $\beta$,
and linear degradation rate $k$.
The glucose dynamics, aside from external forcing (i.e. meals), are driven by
a glucose-dependent (insulin independent) hepatic glucose production $f_\textrm{HGP}$,
an insulin-dependent glucose removal rate $IG$ (with insulin sensitivity factor $s_I$),
and a linear degradation rate $c$.

In this work, we hypothesize that the pancreatic and hepatic regulation of glucose
can instead be approximated by a simple function of glucose $f_\textrm{internal}(G)$.
We also account for the effect of external insulin to the blood glucose level and
add a closure term, $\nu$.
This results in a new single-state equation
\begin{align}
	\label{berg_reduced}
	\dot{G} &= m_\textrm{external}(t) + f_\textrm{internal}(G) + f_\textrm{external}(I) + \nu(t),
\end{align}
where the closure term $\nu(t)$ accounts for additional glycemic dynamics not captured by the first three terms.
To begin evaluating the utility of this perspective, we choose simple forms for these unknown functions.

Specifically, we assume that glucose regulation can be roughly approximated by an exponential decay to a fixed point $G_b$ at rate $\gamma$
such that $f_\textrm{internal}(G) := -\gamma\big(G(t)-G_b\big)$.
We also assume that the effect of external insulin delivery has a simple relationship $f_\textrm{external}(I) := \beta I_\textrm{external}$ with proportionality constant $\beta$.
Finally, we assume that the possibly large residual errors induced by these simplifying assumptions
are given by a Brownian Motion $W(t)$ with variance quantified by $\sigma$.; i.e  $\nu(t):=\sqrt{2\gamma\sigma^2}\dot{W}(t)$. Note that, the term, $\sqrt{2\gamma}$ is included to actually have $\sigma$ to represent the variance of the process and $\gamma$ is a relaxation time-scale. Although counter intuitive in this representation, the solution of the event time model, given in \eqref{disc_general} below, shows the variance of the system dominantly represented by $\sigma$. These choices yield the following Ornstein-Uhlenbeck model for evolution of blood
glucose $G(t)$:
\begin{align}
	\dot{G}(t) &= -\gamma(G(t)-G_b)+m(t)-\beta I(t)+\sqrt{2\gamma\sigma^2}\dot{W}(t).
	\label{cont_general_OU3}
\end{align}

There are four basic parameters for the model in Section \ref{cont_general_OU3}.
$G_b$ (mg/dl) represents the basal glucose (i.e. the mean of the unforced process),
$\gamma$ (1/min) is the decay rate for the exponential mean reversion,
$\beta$ (mg/(dl*U)) is a proportionality constant for the linear effect of IV insulin-based glucose removal,
and $\sigma$ (mg/dl) governs the variance of the oscillations described by $W(t)$.

We use simple models for the meal function $m(t)$
and the insulin delivery function $I(t)$ (defined in Section \ref{event_time_model})
that enable explicit solution of
the continuous time model between events.
We define \emph{events} as times at which the meal or insulin delivery functions change discontinuously, or points at which BG is measured.

The simple linear Gaussian structure of Ornstein-Uhlenbeck models,
along with appropriately simple forcing terms $m(t),I(t)$ (defined in Section \ref{event_time_model}),
allow for tractable solutions to Section \ref{cont_general_OU3}.
Specifically, integration of the system leads to a solution $G(t)$ that is normally distributed with
analytically calculable mean and variance.

\subsection{Advantages of a Linear SDE Model}

In accordance with our goal, which is to develop a highly simplified
yet interpretable model, we work with a forced SDE
of Ornstein-Uhlenbeck type to describe glucose evolution, together
with an observation model of linear form, subject to additive
Gaussian noise.
The Gaussian structure allows for computational
tractability in prediction since probability distributions
on the glucose state are described by Gaussians and hence
represented by simply a mean and variance. With this
modeling choice, we approximate the distribution of
glucose levels by Gaussian distribution. However, this is a
limitation of our modeling approach because the distribution
of glucose levels resembles Gamma distribution rather than
Gaussian distribution. Note that the protocols
for managing glucose depend on intervals; e.g., a goal may be to
keep glucose between 80-150 mg/dl and interval deviation from this
goal, e.g., 151-180 mg/dl, induce changes in the insulin dosage.
This means that decisions are made based on boundaries of glycemic
trajectories.
Nevertheless, because glucose oscillates
under continuous feeding, clinicians typically aim to ensure that the
glycemic mean does not fall below 60 mg/dl or above 180 mg/dl
for any length of time.
The intervals
are then a proxy for this balance of managing the mean and protecting
against trajectories diverging too high or low at any time, including
between observations, \cite{hripcsak2022evaluating}.
Hence accurately
resolving mean and standard deviation in BG levels is important.

The Ornstein-Uhlenbeck process has four contributions: a damping term
which drives the BG level towards its base value at a rate which is possibly
insulin dependent; a forcing representing nutritional intake, exogenously delivered insulin, and a white
noise contribution, which is used to encapsulate the high-frequency
dynamics as these dynamics are difficult to be resolved with sparse
measurements. The presence of noise in the glucose
evolution model, as well as in the data acquisition process, allows for
model error which is natural in view of the the rather simple modeling
framework. Moreover, the existence of the analytic solution of the model
makes it possible to perform parameter estimation with wide range of
filtering and smoothing techniques based on the real-world data.

\subsection{Event-Time Model}\label{event_time_model}

For computational purposes, and because data are typically available at
discrete times, we develop a discrete-time version of the model \eqref{cont_general_OU3}.
We first present it in generality, then develop it specifically for
outpatient Type 2 Diabetes (T2DM) glucose modeling (see Section \ref{event_time_T2D})
and for inpatient intensive care unit (ICU) glucose modeling (see Section \ref{event_time_ICU}).
Note that ICU and T2DM settings are also the focus for our data-driven studies.

The time discretization is defined completely by a dataset in the following sense.
Let $\{t_k^{(m)}\}_{k=1}^{K_m}$ denote the times of relevant nutrition events,
let $\{t_k^{(i)}\}_{k=1}^{K_i}$ denote the times of relevant insulin delivery events,
and let $\{t_k^{(o)}\}_{k=1}^{K_o}$ denote the times of glucose measurements.
We call the re-ordered union of these sets, $$\{t_k\}_{k=0}^N:=\{t_k^{(m)}\}_{k=1}^{K_m}\cup\{t_k^{(i)}\}_{k=1}^{K_i}\cup\{t_k^{(o)}\}_{k=1}^{K_o}$$
as \emph{event times}, where the superscripts, $m$, $i$, and $o$, are used to distinguish the relevant nutrition delivery, insulin delivery, and BG measurement times.

We can obtain the following \emph{event-time model} by
integrating \eqref{cont_general_OU3} over the event-time intervals, $[t_k,t_{k+1})$ for $k=0,1,...,N-1$,
via use of It\^{o} formula, \cite{oksendal2013stochastic}:
\footnote[7]{equivalent to using integrating factors in this case}

\begin{align}
	\begin{aligned}
		G(t_{k+1}) &= G_b+e^{-\gamma h_k}(G(t_k)-G_b)+\int_{t_k}^{t_{k+1}}e^{-\gamma(t_{k+1}-s)}m(s)ds-\int_{t_k}^{t_{k+1}}e^{-\gamma(t_{k+1}-s)}\beta I(s)ds\\
		&\qquad+\sigma\sqrt{1-e^{-2\gamma h_k}}\xi_k,
		\label{disc_general}
	\end{aligned}
\end{align}
where $h_k:=t_{k+1}-t_k$ and $\xi_k\sim N(0,1)$ independent random variables.
We exhibit specific versions of this general event-time model
for T2DM and ICU settings in more detail in the following sections.

\subsubsection{T2DM model}\label{event_time_T2D}

Based on the conditions of T2DM setting detailed in subsection
\ref{clinical_setting}, we set $I(t) \equiv 0$, i.e., we ignore the
exogenous insulin term in the T2DM event-time model. The meal
function, $m(t)$, on the other hand, is essential for capturing
the uptake of glucose into the bloodstream from consumed
carbohydrates. Here, we define $m(t)$ as the difference of
two exponential functions (this choice was shown to be
effective in the T2DM case by \cite{albers2017personalized}):

\begin{equation}
	m(t) = \sum_{k=1}^{K_m}\frac{G_k}{c_k}(e^{-a(t-t_k^{(m)})}-e^{-b(t-t_k^{(m)})})\1_{[t_k^{(m)},\infty)}(t)\label{meal_func_T2D}
\end{equation}

where $t_k^{(m)}$ is the time of the $k^{th}$ meal, $G_k$ (mg/dl) is the total amount of glucose in the $k^{th}$ meal divided by the approximate volume of blood, and  $c_k$ is a dimensionless normalizing constant so that $\frac{1}{c_k}\int_{t_k^{(m)}}^{\infty}(e^{-a(t-t_k^{(m)})}-e^{-b(t-t_k^{(m)})})dt =1$.  Note that $\1_{[a,b)}(\cdot)$ is the indicator function and defined as
	
	\begin{equation*}
		\1_{[a,b)]}(x) = 
		\begin{cases}
			1, & x\in[a,b),\\
			0, & otherwise.
		\end{cases}
\end{equation*}

Therefore, the model in \eqref{cont_general_OU3} becomes
\begin{equation}
	\dot{G}(t) =-\gamma(G(t)-G_b)+\sum_{k=1}^{K_m}\frac{G_k}{c_k}(e^{-a(t-t_k^{(m)})}-e^{-b(t-t_k^{(m)})})\1_{[t_k^{(m)},\infty)}(t)+\sqrt{2\gamma\sigma^2}\dot{W}(t),\label{cont_T2D}
\end{equation}
in the T2DM setting.
In this model, the first term represents the body's own effect to remove insulin from the bloodstream, the second term represents the effect of nutrition on the rate of change of BG, and the last term models the unmodeled dynamics by the first two terms as white noise. Integrating over $[t_0,t]$, we can write the analytic solution of this equation as
\begin{align}
	\begin{aligned}
		G(t)&= G_b+e^{-\gamma(t-t_0)}(G(t_0)-G_b)\\
		&\qquad+\sum_{k=1}^{K_m}\frac{G_k}{c_k}\left(\frac{e^{-a(t-t_k^{(m)})}-e^{-\gamma(t-t_k^{(m)})}}{\gamma-a}-\frac{e^{-b(t-t_k^{(m)})}-e^{-\gamma(t-t_k^{(m)})}}{\gamma-b}\right)\1_{[t_k^{(m)},\infty)}(t)\\
		&\qquad+\int_{t_0}^t e^{-\gamma(t-s)}\sqrt{2\gamma\sigma^2}dW(s).\label{cont_T2D_solt}
	\end{aligned}
\end{align}
Note that, in practice, we need to evaluate BG level at
specific time points and hence need the discrete-time model implied
by the continuous time representation in \eqref{cont_T2D_solt}. Now, by integrating \eqref{cont_T2D} over $[t_k,t_{k+1})$ and denoting $g_k:=G(t_k)$, we obtain
% for $k=0,1,...,N$
%
\begin{equation}
	% g_{k+1} = G_b+e^{-\gamma h_k}(g_k-G_b)+m_k+\sigma\sqrt{1-e^{-2\gamma h_k}}\xi_k, \ \ k=0,1,2,...,N-1,\label{disc_T2D_solt}
	g_{k+1} = G_b+e^{-\gamma h_k}(g_k-G_b)+m_k+\sigma\sqrt{1-e^{-2\gamma h_k}}\xi_k,\label{disc_T2D_solt}
\end{equation}
as a special case of \eqref{disc_general}. Also, for any fixed $t_k$, find the meal times $t_j^{(m)}$ such that $t_j^{(m)}\leq t_k$ and denote the index set of these meal times by $\cI_k$.
Then $m_k$ in \eqref{disc_T2D_solt} becomes
\begin{equation}
	m_k=\sum_{j\in \cI_k}\frac{G_j}{c_j}\left(\frac{e^{-a(t_{k+1}-t_j^{(m)})}-e^{-\gamma h_k}e^{-a(t_k-t_j^{(m)})}}{\gamma-a}-\frac{e^{-b(t_{k+1}-t_j^{(m)})}-e^{-\gamma h_k}e^{-b(t_k-t_j^{(m)})}}{\gamma-b}\right).\label{disc_T2D_meal_func}
\end{equation}
%
% for $k=0,1,2,...,N-1$.
Hence, note that in this case, we have five model parameters to be
estimated: $G_b,\gamma,\sigma,a,b$. Recall that in this setting,
$G_b$ represents the basal glucose value that BG level stays around
starting some time after nutrition intake until the next nutrition intake.
$\gamma$ represents the decay rate of BG level to $G_b$ after the
nutrition intake, and $\sigma$ represents the amplitude of the BG
level oscillations. The parameters $a$ and $b$
entering the meal function implicitly control the time needed for the
glucose nutrition rate to reach its peak value, and the time needed for
this rate to return back to the vicinity of $0$. Because of these
simple physiological meanings, the parameters entering the event-time
model are important not only for accurately capturing, and predicting,
glucose dynamics based on data, but also contain implicit information
about the health condition of the patient. For example, the basal glucose
value is measured during some tests to check if an individual is healthy
pre-diabetic, or diabetic.

\subsubsection{ICU model}\label{event_time_ICU}

The specifics of the ICU setting and the available data, as described
in Section \ref{clinical_setting}, defines the structure of our ICU
model. In this case, we model both the carbohydrate intake, $m(t)$,
and IV insulin delivery, $I(t)$.

We choose to model these external forcings as piecewise constants functions;
this choice corresponds to clinical practice, in which constant infusions are periodically adjusted,
and also allows for simple calculations.

Here, we define the nutritional forcing function as

\begin{equation}
	m(t) = \sum_{k=1}^{K_m}  d_k \1_{[t_k^{(m)},t_{k+1}^{(m)})}(t),\label{meal_func_ICU}
\end{equation}

where $t_k^{(m)}$ is the time at which a clinician changes the nutrition delivery rate,
$d_k$ is the nutrition rate over the time interval $[t_k^{(m)},t_{k+1}^{(m)})$;
these features are both directly available in our clinical dataset.

Similarly, we define the external insulin delivery rate as

\begin{equation}
	I(t) = \sum_{k=1}^{K_i} i_k\1_{[t_k^{(i)},t_{k+1}^{(i)})}(t),\label{insulin_func_ICU}
\end{equation}

where $i_k$ is the rate of insulin over the time interval $[t_k^{(i)},t_{k+1}^{(i)})$, again obtained directly from the dataset.

Therefore, substituting \eqref{meal_func_ICU} and \eqref{insulin_func_ICU} into the general equation \eqref{cont_general_OU3}, the ICU version of our model becomes

\begin{align}
	\dot{G}(t) = -\gamma(G(t)-G_b)+\sum_{k=1}^{K_m}  d_k \1_{[t_k^{(m)},t_{k+1}^{(m)})}(t)-\beta \sum_{k=1}^{K_i} i_k \1_{[t_k^{(i)},t_{k+1}^{(i)})}(t)+\sqrt{2\gamma\sigma^2}\dot{W}(t).
	\label{cont_ICU}
\end{align}

In this model, the first term models the glucose removal rate with the body's own effort ($\gamma$),
the second term shows the effect of nutrition $m(t)$ on the BG level,
the third term, $\beta I(t)$, models the external insulin effect,
and the last term models unmodeled dynamics by the first three terms as a white noise term.

We integrate \eqref{cont_ICU} to get the analytical solution for any $t\geq t_0$ as follows

\begin{align}
	\begin{aligned}
		G(t) &= G_b + e^{-\gamma(t-t_0)}(G(t_0)-G_b)+ \sum_{k=1}^{K_m} d_k\int_{t_0}^t e^{-\gamma(t-s)}\1_{[t_k^{(m)},t_{k+1}^{(m)})}(s)ds\\
		&\qquad -\beta\sum_{k=1}^{K_i} i_k\int_{t_0}^t e^{-\gamma(t-s)}\1_{[t_k^{(i)},t_{k+1}^{(i)})}(s)ds+ \sqrt{2\gamma\sigma^2} \int_{t_0}^t e^{-\gamma(t-s)} dW(s).
		\label{cont_ICU_solt}
	\end{aligned}
\end{align}

As in the previous section, we can also integrate \eqref{cont_ICU} over $[t_k,t_{k+1})$ to obtain solutions at event-times, with $g_k = G(t_k)$,

\begin{align}
	\begin{aligned}
		g_{k+1} &= G_b + e^{-\gamma h_k}(g_k-G_b) + \frac{1}{\gamma}(1-e^{-\gamma h_k})d_k - \beta\frac{1}{\gamma}(1-e^{-\gamma h_k})i_k+ \sigma\sqrt{1-e^{-2\gamma h_k}}\xi_k
		% &\qquad \ \ k=0,1,2,...,N-1,
		\label{disc_ICU_solt}
	\end{aligned}
\end{align}

as another special case of \eqref{disc_general}.
Here, we have four model parameters to estimate: $G_b,\gamma,\sigma,\beta$.
Remember once again, $G_b$ is the basal glucose value and $\gamma$ is the decay rate of the BG level to its basal value, and $\sigma$ is a measure for the magnitude of the BG oscillations.
Finally, $\beta$ is a proportionality constant, which is used to scale the effect of IV insulin on the BG rate change appropriately. These four parameters represent physiologically valid quantities that could properly resolve the mean and variance of the BG level.

\subsection{Parameter Estimation}\label{param_est}

In this section we formulate the parameter estimation
problem. %In \Cref{bayesian_form}
We construct an overarching Bayesian framework for our
parameter estimation problems. We then describe two solution approaches
for this problem: an  optimization based approach which identifies the
most likely solution, given our model and data assumptions; and Markov Chain Monte Carlo (MCMC), which
samples the distribution on parameters, given data, under the same  model
and data assumptions. These two solution approaches are detailed in the Supplementary Material.

As shown in detail before, our model takes slightly different forms in the T2DM and ICU settings. In the former the model parameters to be estimated are $G_b,\gamma,\sigma,a,b$ whereas in the latter the unknown parameters are $G_b,\gamma,\sigma,\beta$. However, we adopt a single approach to parameter estimation. To
describe this approach we let the vector, $\theta$ represent the unknown model
parameters to be determined from the data, noting that this is a different set
of parameters in each case.
Many problems in biomedicine, and the problems we study here in particular,
have both noisy models and noisy data, leading to a relationship between
parameter $\theta$ and data $y$ of the form
\begin{equation}
	\label{eq:yth}
	y=\mathcal{G}(\theta,\zeta)
\end{equation}
where unknown $\zeta$ is a realization of a mean zero random variable, but
its value is not known to us. The objective is to
recover $\theta$ from $y$. We will show how our model of the glucose regulatory
system lead to such a model.

The Bayesian approach to parameter estimation is desirable for two primary reasons: first it allows for seamless incorporation of imprecise prior information with uncertain mathematical model and noisy data, by adopting a formulation in which all variables have probabilities associated to them; secondly it allows for the quantification of uncertainty in the parameter estimation. Whilst extraction of information from the posterior probability distribution on parameters given data is challenging, stable and practical computational methodology based around the Bayesian formulation has emerged over the last few decades;  see \cite{stuart2010inverse}. In this work, we will follow two approaches: (a) obtaining the
{\em  maximum a posteriori (MAP) estimator}, which leads to an optimization
problem for the most likely parameter given the data, and (b) obtaining
\emph{samples} from the posterior distribution on parameter given data,
using MCMC techniques.

Now let us formulate the parameter estimation problem. Within the event-time
framework, let $g=[g_k]_{k=0}^N$ be the vector of BG levels at event times $\{t_k\}_{k=0}^N$, and $y=[y_k]_{k=1}^{K_o}$ be the vector of measurements at the measurement times $\{t_k^{(o)}\}_{k=1}^{K_o}\subset\{t_k\}_{k=0}^N$. By using the event-time version, and defining $\{\xi_k\}_{k=0}^N$ to be independent and identically distributed standard normal random variables, we see that given the parameters $\theta$, $g$ has multivariate normal distribution, i.e., $\mathbb{P}(g|\theta)=N(m(\theta),C(\theta))$. Equivalently,

\begin{equation}
	g = m(\theta)+\sqrt{C(\theta)}\xi, \ \ \xi\sim N(0,I).\label{vector_form_g}
\end{equation}

Let $L$ be a $K_o\times(N+1)$ matrix that maps $\{g_k\}_{k=0}^N$ to $\{y_k\}_{k=1}^{K_o}$. That is, if a measurement $i\in{1,...,K_o}$ is taken at the event time $t_j$, $j\in{0,1,...,N}$, then the $i^{th}$ row of $L$ has all 0's except the $(j+1)^{st}$ element, which is 1. Adding a measurement noise, we state the observation equation as follows:

\begin{equation}
	y = Lg+\sqrt{\Gamma(\theta)}\eta, \ \ \eta\sim N(0,I),\label{vector_form_y_1}
\end{equation}

where $\Gamma(\theta)$ is a diagonal matrix representing the measurement noise. Thus, we obtain the likelihood of the data, given the glucose time-series and the parameters, namely
$$\mathbb{P}(y|g,\theta)=N(Lg,\Gamma(\theta)).$$
However, when performing parameter estimation, we are not
interested in the glucose time-series itself, but only in the parameters.
Thus we directly find the likelihood of the data given the parameters
(implicitly integrating out $g$)
by combining \eqref{vector_form_g} and \eqref{vector_form_y_1} to obtain

\begin{equation}
	y = Lm(\theta)+\sqrt{S(\theta)}\zeta, \ \ \zeta\sim N(0,I),\label{vector_form_y_2}
\end{equation}

where $S(\theta)=LC(\theta)L^T+\Gamma(\theta)$. Since $\zeta$ has multivariate normal distribution, using the properties of this distribution, we find that given the parameters, $\theta$, $y$ also has multivariate normal distribution with mean $Lm(\theta)$ and covariance matrix $S(\theta)$.
This is the specific instance of equation \eqref{eq:yth} that arises
for the models in this paper.

We have thus obtained $\mathbb{P}(y|\theta)=N(Lm(\theta),S(\theta))$, that is,

\begin{equation}
	\mathbb{P}(y|\theta) = \frac{1}{\sqrt{(2\pi)^{K_m}\det(S(\theta)))}}\exp\left(-\frac{1}{2}(y-Lm(\theta))^TS(\theta)^{-1/2}(y-Lm(\theta))\right)\label{likelihood};
\end{equation}

this is the likelihood of the data, $y$, given the parameters, $\theta$. Also, since we prefer to use $-\log(\mathbb{P}(y|\theta))$ rather than directly using $\mathbb{P}(y|\theta)$ for the sake of computation, we state it explicitly as follows:

\begin{equation}
	-\log(\mathbb{P}(y|\theta)) = \frac{K_m}{2}\log(2\pi)+\frac{1}{2}\log(\det(S(\theta)))+\frac{1}{2}(y-Lm(\theta))^TS(\theta)^{-1}(y-Lm(\theta)).\label{log_likelihood}
\end{equation}

Moreover, by using Bayes Theorem, we write

\begin{equation}
	\mathbb{P}(\theta|y)=\frac{\mathbb{P}(y|\theta)\mathbb{P}(\theta)}{\mathbb{P}(y)}\propto\mathbb{P}(y|\theta)\mathbb{P}(\theta).\label{bayes}
\end{equation}

Note that the second statement of proportionality
follows from the fact that the term, $\mathbb{P}(y)$, on the denominator is constant with respect to the parameters, $\theta$, and plays the role of a normalizing constant.

From another point of view, considering \eqref{vector_form_g} and \eqref{vector_form_y_2}, we see that given $\theta$, $(g,y)$ has multivariate normal distribution with mean and covariance matrix that could be computed from the above equations since, given $\theta$, everything is explicitly known. Then, integrating $g$ out, in other words, computing the marginal distribution we obtain the distribution of $y|\theta$, which corresponds to the one stated in \eqref{vector_form_y_2}.

Now, to define the prior distribution $\mathbb{P}(\theta)$ we assume that the unknown parameters are distributed uniformly across a bounded set $\Theta$ and define

\begin{equation}
	\mathbb{P}(\theta)=\frac{1}{|\Theta|}\1_{\Theta}(\theta)=
	\begin{cases}
		\frac{1}{|\Theta|}, \ \ &\theta\in\Theta,\\
		\hfil 0, \ \ &\theta\notin\Theta,
	\end{cases}\label{prior_dist}
\end{equation}

where $\1_{\Theta}(\cdot)$ is the indicator function and $|\Theta|$ is the volume of the region defined by $\Theta$. Thus, by substituting the likelihood, \eqref{likelihood}, and the prior distribution, \eqref{prior_dist}, into \eqref{bayes}, we formulate the posterior distribution as follows
\begin{equation}
	\mathbb{P}(\theta|y) \propto \frac{1}{|\Theta|\sqrt{(2\pi)^{K_m}\det(S(\theta)))}}\exp\left(-\frac{1}{2}(y-Lm(\theta))^TS(\theta)^{-1/2}(y-Lm(\theta))\right)\1_{\Theta}(\theta).\label{posterior_pdf}
\end{equation}

Then, we use this posterior distribution to state the parameter estimation problem whose details can be found in the Supplementary Material.

\subsection{Experimental Design}\label{exp_design}

In this section, we describe the datasets in more detail, the experiments that we design to present our results, and the methods that we follow to perform parameter estimation and forecasting. Depending on the specifics of each case and to reflect the real-life situation, we designed different experiments in the T2DM and ICU settings. However, the mathematical approaches for parameter estimation and forecasting stay the same for both settings because we use similar mechanistic models.

We theoretically define the observational noise covariance $\Gamma(\theta)$, given in
\eqref{vector_form_y_1}, to be a diagonal matrix with form
$diag(\Gamma(\theta)):=\lambda*Lm(\theta)$, which represents that it is proportional to the mean BG level. However, we observed that the variation in glycemic response, which we will define later more formally, is the sum of the measurement noise and personal glycemic variation, accounted by the model parameter, $\sigma$. Because of this relationship, for more accurate estimation of $\sigma$, we set the measurement noise to 0. Note that this is only a practical choice and with this choice, we can still estimate the variation in glycemic response accurately.

\subsubsection{T2DM}\label{T2D_model_params_forc}

\paragraph{Model, Parameters, and Dataset}

In this setting, we use the model \eqref{disc_T2D_solt} with the function $m_k$ defined as in \eqref{disc_T2D_meal_func}. Hence, there are five parameters to be estimated: basal glucose value, $G_b$, BG decay rate $\gamma$, the measure for the amplitude of BG oscillations, $\sigma$, and $a$ and $b$, which are the parameters implicitly modeling the time needed for the rate of glucose in the nutrition entering the bloodstream to reach its maximum value and the total time
needed for this rate to decrease back to 0. We assume that the prior distribution is non-informative and initially the parameters are independent,
except for a constraint on the ordering of $a$ and $b$. We determine realistic lower and upper bound values for each of them, define $\Theta':=[0,750]\times[0.01,0.5]\times[0,100]\times[0.01,0.05]\times[0.01,0.05]$ (in the order of $G_b,\gamma,\sigma,a,b$), and then
define $\Theta$ from $\Theta'$ by adding the constraint $a<b.$  We thereby
form the prior distribution as defined in \eqref{prior_dist}. Recall that
these bounds define the constraints employed when we define the parameter
estimation problem in the optimization setting for the MAP point. The
set $\Theta$ is determined from clinical and physiological prior knowledge,
and by simulating the model \eqref{cont_T2D}
and requiring realistic BG levels. Data are collected from three different T2DM patients. Detailed information on the dataset can be found in Table \ref{tab1}.

\begin{table}[!ht]
	\centering
	\begin{tabular}{|l|c|c|c|}
		\hline
		Patient ID			&	patient 1	&	patient 2	&	patient 3	\\ \hline
		Total \# glucose measurement	&	304	&	211	&	91			 \\ \hline
		Total \# meals recorded	&	122 	&	76	&	46			\\ \hline
		Total \# days measured	&	26.6	&	27.67	&	28.12		\\ \hline
		Mean measured glucose &	113$\pm$25	&	127$\pm$32	&	124$\pm$26	\\ \hline
		Training set: \# glucose measurement	&	80	&	53	&	29		\\ \hline
		Training set: \# meals recorded	&	26	&	18	&	15			\\ \hline
		Training set: \# days measured	&	7.02	&	7	&	7.05		\\ \hline
		Training set: mean measured glucose	&	112$\pm$25	&	116$\pm$28	&	125$\pm$24	\\ \hline
		Testing set: \# glucose measurement	&	224	&	158	&	31		\\ \hline
		Testing set: \# meals recorded	&	96	&	58	&	62		\\ \hline
		Testing set: \# days measured	&	19.58	&	20.67	&	21.07		\\ \hline
		Testing set: mean measured glucose	&	113$\pm$25	&	130$\pm$33	&	123$\pm$27	\\ \hline
	\end{tabular}
	\caption[T2DM dataset]{Information about the dataset that is used in the T2DM setting, which is collected from three different T2DM patients. Note that there is a considerable variability between the data collection behavior of each patient, which is also reflected in the number of recorded measurements and meals. Also, recall that we intentionally used one week of data for training and the following three week of data for testing.}
	\label{tab1}
\end{table}

\paragraph{Parameter Estimation and Uncertainty Quantification}\label{pe_uq_T2D}

We perform parameter estimation for three patients separately. First, we estimate parameters by using data over four consecutive, non-overlapping time intervals with optimization and MCMC approaches. Besides estimated values, we also provide UQ results. In the optimization setting, we use
the Laplace approximation as detailed in the Supplementary Material.
The optimal parameters determine the mean of the Gaussian approximation,
and the inverse of the Hessian matrix becomes the covariance matrix, providing the tools for UQ. In the MCMC approach, we use the resulting random samples for UQ.

\paragraph{Forecasting}\label{forc_T2D}

\begin{figure}[ht]
	\centering
	\includegraphics[scale=0.2]{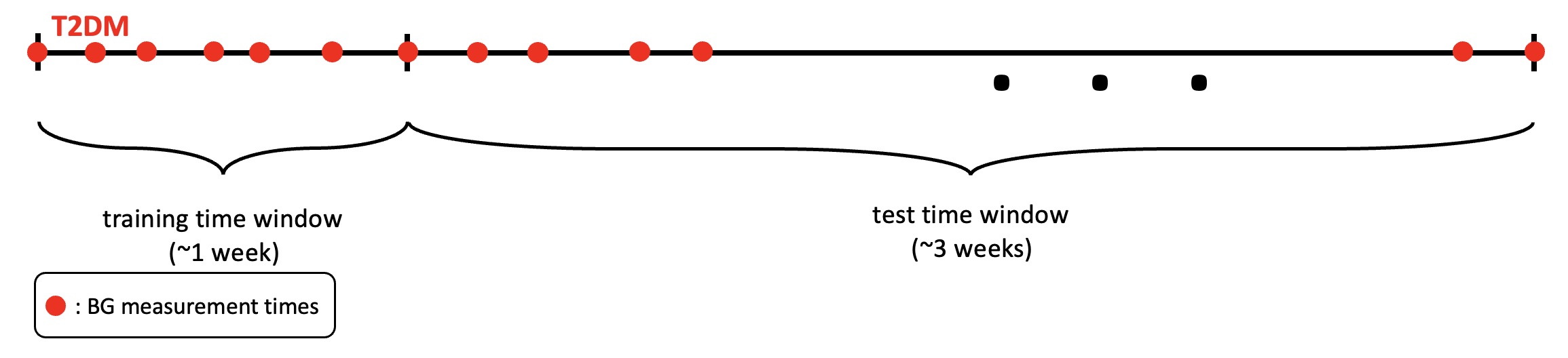}
	\caption{This schema shows how we divided T2DM patients' data into training and test time windows. For each patient, the first week of data is the training time window used to estimate the model. The estimated model represents each patient's personalized BG behavior and is used to forecast BG values over the test time window, which is of length three weeks and follows the training time window.}
	\label{fig1}
\end{figure}

We adopt a train-test set-up as follows. Since the health
conditions of the T2DM patients are unlikely to change
over time intervals that are on the order of days, we design
an experiment in which we use one week of data for
estimating the patient-specific parameters. Then, we use
the estimated parameters to form a patient-specific model
and use this model to forecast BG levels for the following three weeks, using the known
glucose input through the meals; this leads to a three-week testing
phase. We provide a visual representation of this process in Fig~\ref{fig1}.
From a practical patient-centric point of view this leads to
a setting in which forecasting BG levels for the following three weeks
requires patients to collect BG data for only one week in every month,
and then the patient-specific model will be able to capture their dynamics
and provide forecasts based on nutrition intake data over the rest of the
month.

\subsubsection{ICU}\label{sec:icu_exp}

\paragraph{Model, Parameters, and Dataset}

In the ICU setting, we use the model \eqref{disc_ICU_solt}, and there are now four parameters to be estimated: basal glucose value, $G_b$, BG decay rate, $\gamma$, the parameter used to quantify the amplitude of the oscillations in the BG level, $\sigma$, and a proportionality constant, $\beta$ to scale the effect of insulin IV on the BG level. Similar to what we did in the T2DM setting, we find realistic lower and upper bounds for the unknown parameter values and set $\Theta:=[0,750]\times[0.02,0.5]\times[0,100]\times[20,110]$ to obtain the prior distribution as defined in \eqref{prior_dist}. In this case, we impose two further linear constraints, namely $G_b-3.5*\beta<115$ and $\beta-1110\gamma<10$. These constraints
are imposed to ensure that the model predictions remain biophysically
plausible, and are determined simply by forward simulation of the SDE model;
the resulting inequality constraints do not overly constrain the
parameters in that good fits can be found which satisfy these constraints,
and yet they yield more realistic BG level behavior than solutions
found without them. Thus as in the T2DM case, we choose the bounds and the constraints based on physiological knowledge and requiring simulated BG levels resulting from values within the region $\Theta$ to be realistic.

\begin{table}[!h]
	\centering
	\begin{tabular}{|l|c|c|c|}
		\hline
		Patient ID			&	patient 4	&	patient 5	&	patient 6	\\ \hline
		Total \# glucose measurement	&	177	&	204	&	271			 \\ \hline
		Total \# days measured	&	13.99	&	16.8	&	24.48		\\ \hline
		Mean measured glucose &	141$\pm$18	&	151$\pm$32	&	151$\pm$43	\\ \hline
		Training set: average \# glucose measurement &	14.13	&	13.5	&	14.07		\\ \hline
		Testing set: average \# glucose measurement	&	1	&	1	&	1		\\ \hline
	\end{tabular}
	\caption[ICU dataset]{Information about the dataset that is used in the ICU setting, collected from three ICU patients who are not T2DM. Because of the experiment we designed the training sets are moving with by overlapping with each other. So, we provide average number of glucose measurements over these moving windows. Also, since we forecast until the next measurement time following the training time window, each testing set contains only one glucose measurement. Other information that is included in Table \ref{tab1}, but not here, such as mean measured glucose over training set(s) is neither meaningful nor helpful in this setting.}
	\label{tab2}
\end{table}

Summary statistics about our ICU dataset can be found in Table \ref{tab2}. Note that in this case, we used all available data for each patient to perform parameter estimation and forecasting, and all three ICU patients are non-T2DM.

\paragraph{Parameter Estimation and Uncertainty Quantification}\label{pe_uq_ICU}

We use both the optimization and MCMC approaches for parameter estimation in a patient-specific manner, in this setting, too. However, for UQ, we use only MCMC to estimate the posterior mean and
variance on the parameter; this is because there were cases where
it was not appropriate to use the Laplace approximation, something
that will be explained in more detail in Section \ref{num_res_ICU}.

\begin{figure}[!b]
	\centering
	\includegraphics[scale=0.2]{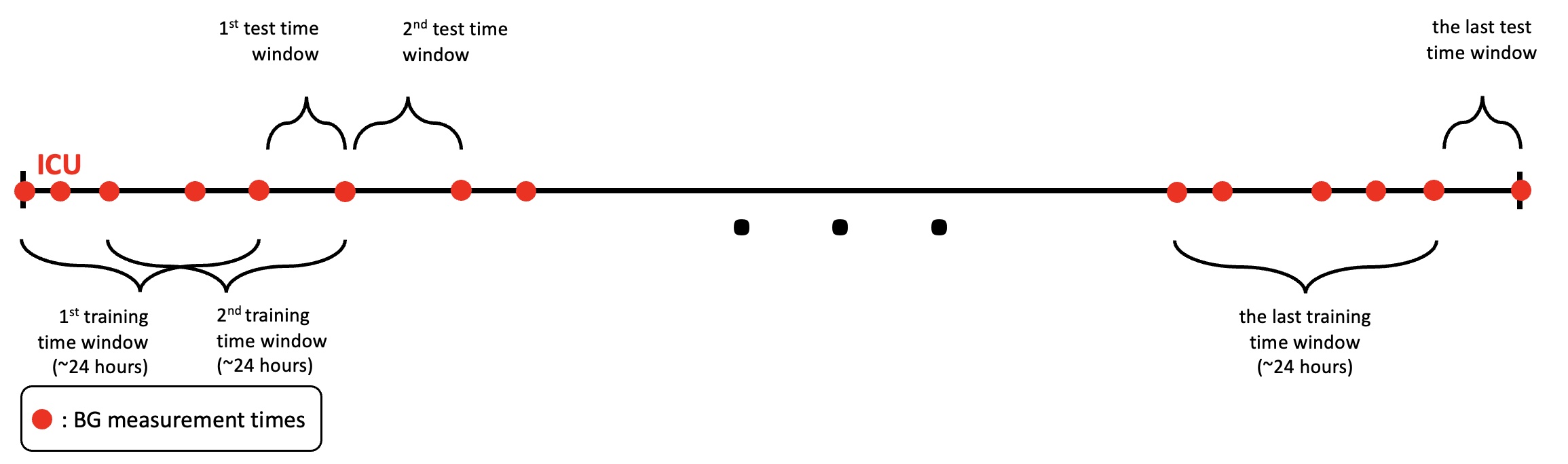}
	\caption{This schema shows our experimental design for the ICU setting. Each training time window has a length of approximately 24 hours, which is used for model estimation. Then, we forecast the first BG value measured after the training time window. We perform this prediction for the whole dataset by moving the time windows.}
	\label{fig2}
\end{figure}

\paragraph{Forecasting}\label{forc_ICU}

The characteristics of the health conditions of ICU patients are described
in Section \ref{clinical_setting}. The abrupt changes in their health
conditions are reflected in the model parameters. To avoid compensating
for different values of parameters over longer time intervals, and to make
more accurate predictions, we use only one day of data for parameter
estimation in the ICU. Moreover, to construct an experiment that reflects
real-life scenarios, we need be able to estimate the model parameters with
smaller size datasets than in the T2DM case, because of the imperative
of regular intervention within the ICU setting, typically on a time-scale
of hours. As a consequence our train-test set-up in this case differs
quantitatively from the T2DM case. The training sets for each patient
consist of approximately one day of data over a moving time intervals,
with end points chosen to be BG measurement times. Thus, the time
windows are obtained by moving the right end point to the next BG
measurement time and choosing its left end point with the constraint
that it contains approximately one day of data. In this case, there is a
large overlap between the consecutive time windows of the training sets.

On the other hand, because of rapidly changing conditions,
forecast of BG levels needs only to be accurate over shorter
time-scales. It is important to know glycemic dynamics on
the order of hours (not days) to manage the glycemic response of ICU
patients. Thus, the test time windows include only one BG measurement,
which is the next BG measurement collected right after the BG measurement
defining the right end point of the corresponding training time window. We
follow the same procedure over the moving time intervals to the end of the
whole dataset for each patient. We visually exhibit this procedure in
Fig~\ref{fig2}. From a practical point of view,
this experiment exhibits a real life situation in which we use only one day
of data for parameter estimation and then perform forecasting for the
next few hours based on the estimated parameters. Such a set-up
would be desirable as a support to glycemic management of these patients.

\subsection{Model Evaluation}

In this section, we introduce the statistics that we will use to evaluate and
compare the forecasting capability of the models. Let $\{y_i\}_{i=1}^N$ denote the true BG measurements over the predefined testing time window for an experiment. Let $\{\hat{y}_i\}_{i=1}^N$ denote the forecast obtained by a model at the measurement time points . Note that for a stochastic model, $\{\hat{y}_i\}_{i=1}^N$ represents the mean of the model output. When a stochastic model is used, it is natural to obtain a confidence interval as this may be obtained as a direct consequence of the fact that the model output is in the form of a random variable; such an output cannot be obtained for an ODE model
when parameters are learned through optimization. However, by using appropriate parameter and state estimation techniques, it may again be possible to obtain a
similar kind of confidence interval for the model output which is in the form of a point-estimate. When we have probabilistic forecasts we let
$\{\epsilon_i\}_{i=1}^N$ denote the corresponding standard deviation for each forecast at the true measurement points so that we can form 1- and 2-stdev bands as $[\hat{y}_i-\epsilon_i,\hat{y}_i+\epsilon_i]_{i=1}^N$ and $[\hat{y}_i-2\epsilon_i,\hat{y}_i+2\epsilon_i]_{i=1}^N$, respectively. Then, for each model, we can compute the percentage of true measurements, $\{y_i\}_{i=1}^N$, that are captured in their respective 1- and 2-stdev bands. These percentages will be the tools that we will use for evaluation. In addition, we will use standard measures such as root-mean-squared error (RMSE), mean percentage error (MPE), and Pearson's correlation coefficient, (CORR) which are computed as follows.

\begin{equation*}
	RMSE = \sqrt{\frac{1}{N}\sum_{i=1}^N (y_i-\hat{y}_i)^2}, \ \ \ \ MPE = \sum_{i=1}^N \frac{|y_i-\hat{y}_i|}{y_i}*100, \ \ \ \ CORR = corr\left(\{y_i\}_{i=1}^N,\{\hat{y}_i\}_{i=1}^N\right)
\end{equation*}

In addition to these metrics, we compare the forecasting accuracy of this model with other physiology-based mechanistic models. In T2DM setting, we use the longitudinal diabetes pathogenesis model (LDP), \cite{ha2020type} , describing BG dynamics of T2DM patients. In the ICU setting, we use ICU Minimal Model (ICUMM), introduced in	\cite{vanherpe06,vanherpe07} describing BG dynamics of ICU patients. Also, in both settings, we use simply the mean and variance computed from the respective training data for comparison. We call this model, mean-variance model. We provide more detail about these models in Sections \ref{t2dm_comparison} and \ref{icu_comparison}.
	
Here, by \textit{comparing models}, we mean comparing their forecasting accuracy. However, both the LDP model and ICUMM are nonlinear mechanistic models while our model is a linear mechanistic model. Using these models within predictive algorithms requires computational model estimation to solve for the best model parameters and states that fit the patient data. A model estimation problem formulated based on a nonlinear model has multiple minima while the one formulated based on a linear model generally has a unique minimum that produces the optimal solution to model estimation problem. However, when there is multiple minima, it is almost impossible to find estimate the global minimum. Because of these characteristic differences, an absolute comparison between a nonlinear model and a linear model is impossible. Therefore, comparison of prediction accuracy between these different type of models should be carefully handled. For example, obtaining a smaller error with a linear model does not imply that this model is \textit{better} than the nonlinear model, as it is unknown if the global minimum is reached by the nonlinear model. However, comparing the prediction accuracy is useful to have a sense of the level of prediction accuracy achieved by these models.

\section{Results}\label{num_results}

In this section, we present results concerning the simple yet interpretable
model introduced in this paper; we now refer to this as the minimal stochastic glucose
(MSG) model. The two primary conclusions are that:
\begin{itemize}
	
	\item We obtain BG forecasting results at least as accurate as other established
	models in both the T2DM and ICU settings, \cite{vanherpe07}, and the uncertainty bands with which
	we equip our forecasts play an important role in this regard;
	
	\item We learn a substantial amount about the interpretable parameters within the
	models, with possible clinical uses deriving from the parameter estimates, and from
	tracking them over time, again using the uncertainty measures that accompany
	them as measures of confidence.
	
\end{itemize}
The combination of simple predictive model and data
acquisition accounts for the uncontrolled and complex nature of the data, including data sparsity,
inaccuracy, noisiness, non-stationarity, and biases resulting from the health care process \cite{albers2018estimating,albers2010statistical,hripcsak2013correlating,hripcsak2017high,hripcsak2012next,hripcsak2011exploiting,levine2018methodological,pivovarov2014identifying},
whilst also being interpretable and leading to patient-specific parameter inference and prediction. Even though the MSG model is relatively simple it is not always identifiable, given data. For example, having two parameters, $\gamma$ and $\beta$, related to BG decay rate in the ICU context made it hard to identify these parameters accurately because of the complexities mentioned above. Despite lack of identifiability of some parameters, parameters as estimated lead to models which are able to forecast and represent the glucose dynamics. To answer whether the parameter estimates,
forecasts, and uncertainty quantification \emph{are} good enough to impact clinical
understanding and decision-making or to construct physiologically-anchored phenotypes
would require evaluation, \cite{albers2012population,albers2014dynamical,albers2018mechanistic}
e.g., manual chart review in conjunction with a qualitative trial of clinical decision-making
or a phenotyping analysis respectively. In the absence of these analyses we will rely on
face validity, \cite{churchill1979paradigm,hardesty2004use,weiner2010corsini}, to evaluate
effectiveness of the model in representing the dynamics and in forecasting.

\subsection{T2DM}\label{num_res_T2D}

In this setting, our results demonstrate the
effectiveness of the MSG model in capturing the patients’ BG dynamics. Specifically
the effectiveness is reflected in the estimated parameter values and in forecasting
future BG levels, using these parameters, over time periods of length up to three weeks.

\subsubsection{Parameter Estimation}

Our results exhibit three substantive pieces of evidence that support the validity of the model and its potential effectiveness for understanding the physiologic state of an individual and forecasting. \emph{First}, the estimated model parameter values and their evolution over time are physiologically valid. That is, the estimated values reflect the patient's state as evaluated given available data. Moreover, the evolution of the estimated parameter values over time reflects changes in the patients' states in a manner consistent with both the data and what is known about the non-stationary nature of T2DM. \emph{Second}, the UQ intervals for the estimated parameters are physiologically plausible and have three features that make the model potentially useful: (i) relative to the value of the estimated parameter, the UQ intervals are wide enough to provide information on the reliability of the point estimates, (ii) the UQ intervals' evolution over time, demonstrating sensitivity to time and the ability to adapt to non-stationary patients, and (iii) the UQ intervals are narrow enough to plausibly be used to differentiate behavior choices, such as carbohydrate consumption. And \emph{third}, the UQ and parameter estimation appears to be robust; different estimation methods arrive at similar results. A comparison of the estimated parameter values and corresponding UQ intervals obtained using optimization and MCMC are very similar in almost all of the cases, supporting the robustness of the estimates and relative insensitivity to the estimation methodology. Together, these features imply that with a reasonable inference scheme, this model could provide useful information for decision-making and a robust clinical understanding of the patient.

\begin{figure}[!ht]
	\begin{subfigure}{0.33\textwidth}
		\includegraphics[scale=0.14]{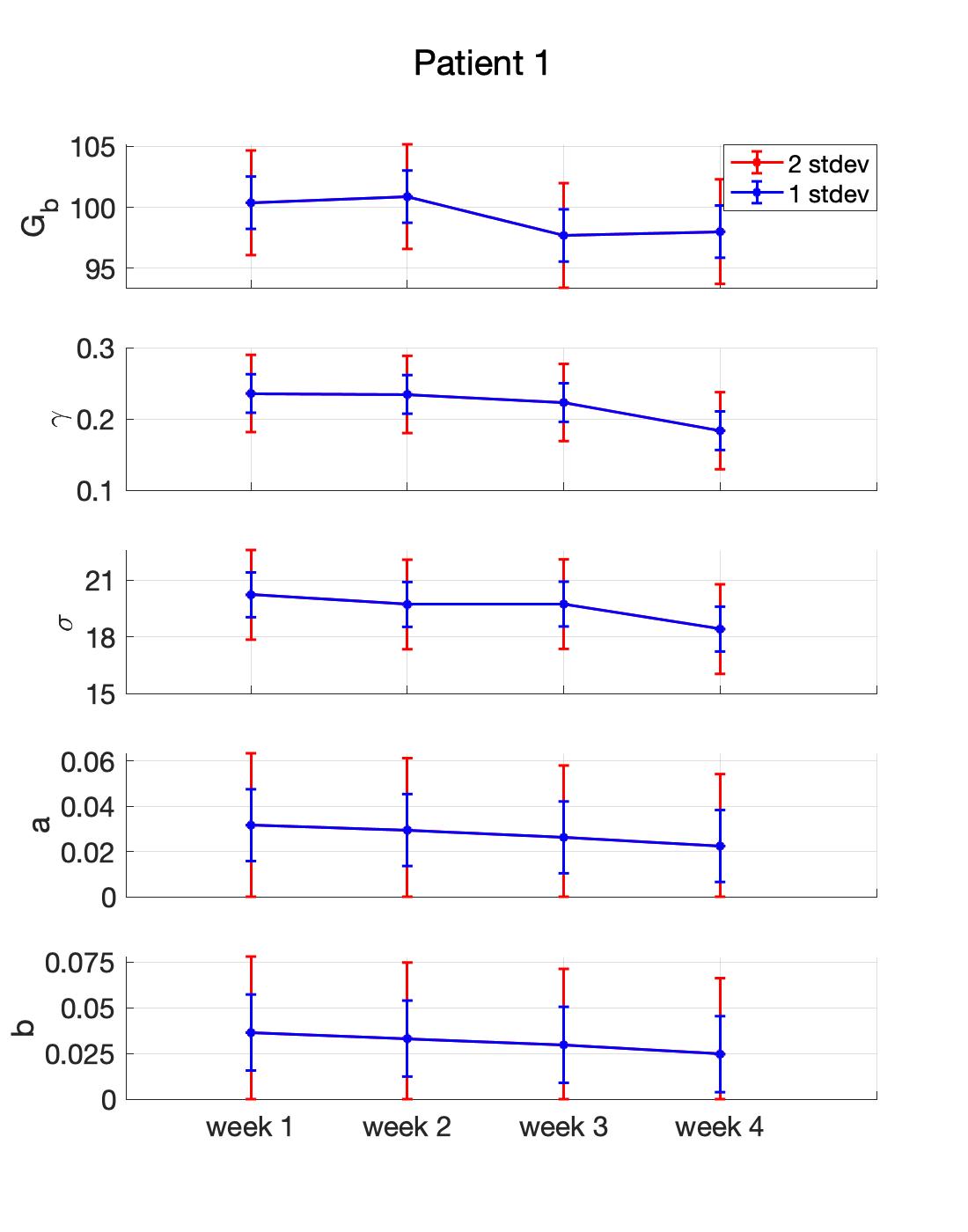}
		\caption{optimization} \label{fig3A}
	\end{subfigure}\hspace{0cm}
	\begin{subfigure}{0.33\textwidth}
		\includegraphics[scale=0.14]{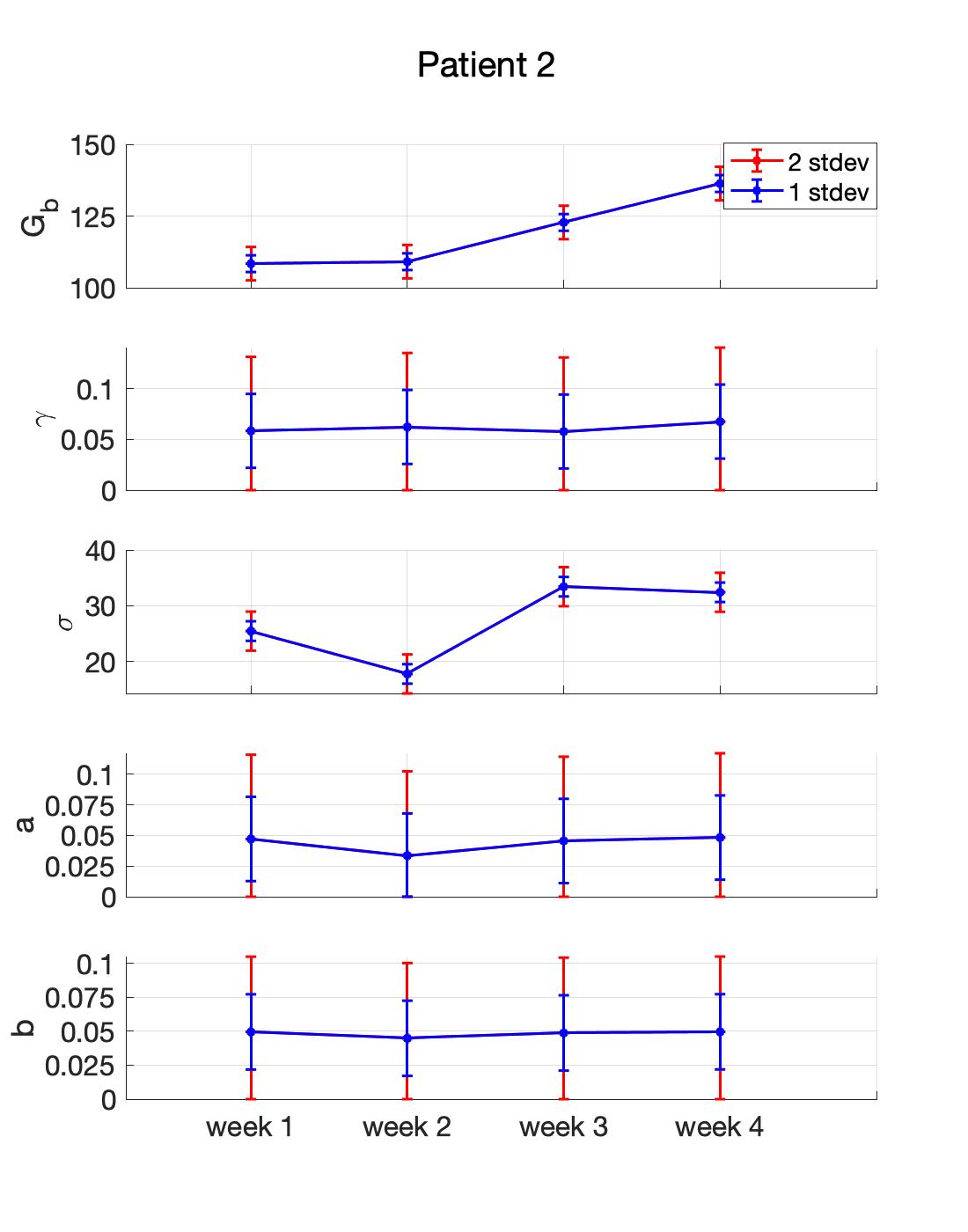}
		\caption{optimization} \label{fig3B}
	\end{subfigure}\hspace{0cm}
	\begin{subfigure}{0.33\textwidth}
		\includegraphics[scale=0.14]{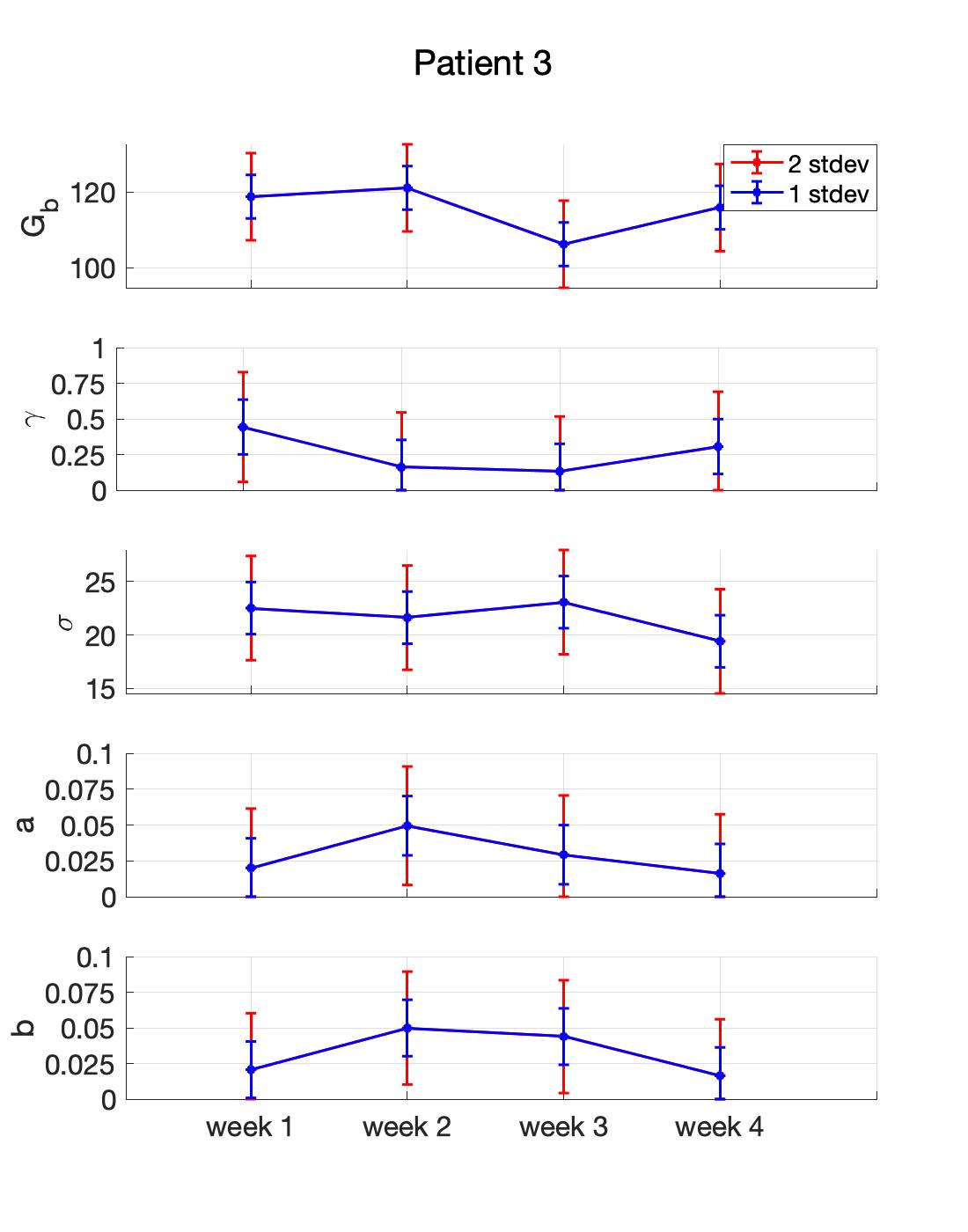}
		\caption{optimization} \label{fig3C}
	\end{subfigure}\hspace{0cm}
	
	\medskip
	
	\begin{subfigure}{0.33\textwidth}
		\includegraphics[scale=0.14]{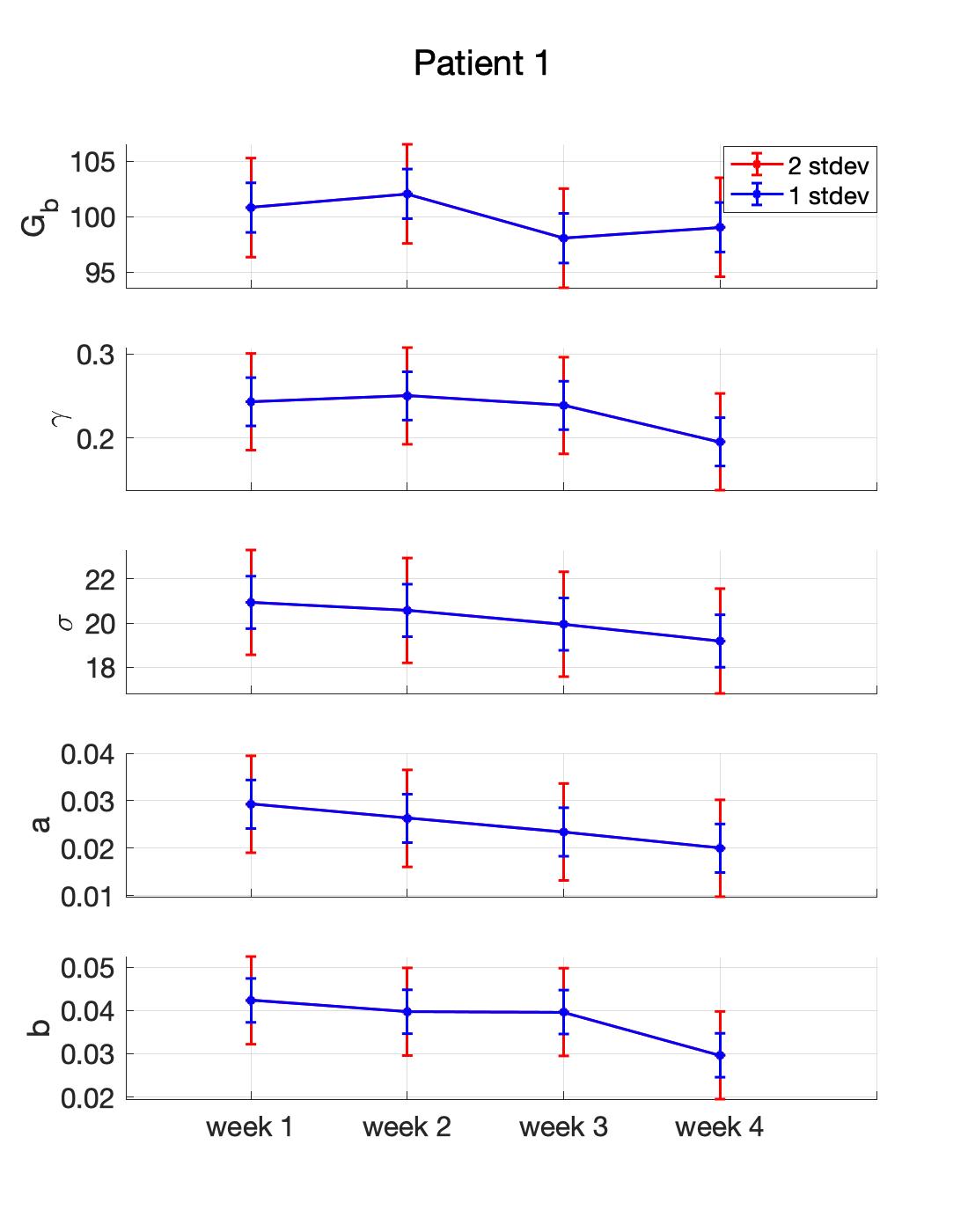}
		\caption{MCMC} \label{fig3D}
	\end{subfigure}
	\begin{subfigure}{0.33\textwidth}
		\includegraphics[scale=0.14]{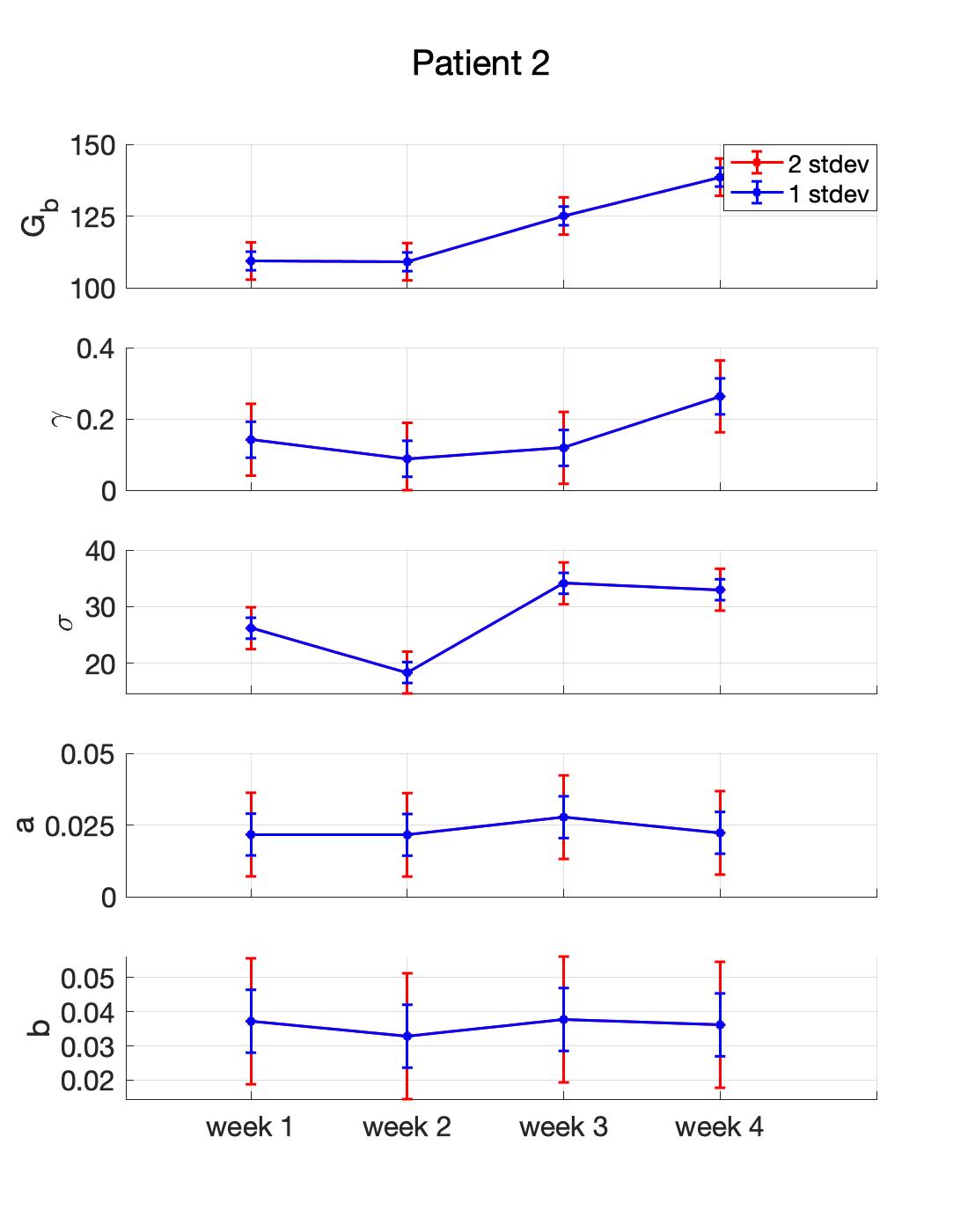}
		\caption{MCMC} \label{fig3E}
	\end{subfigure}
	\begin{subfigure}{0.33\textwidth}
		\includegraphics[scale=0.14]{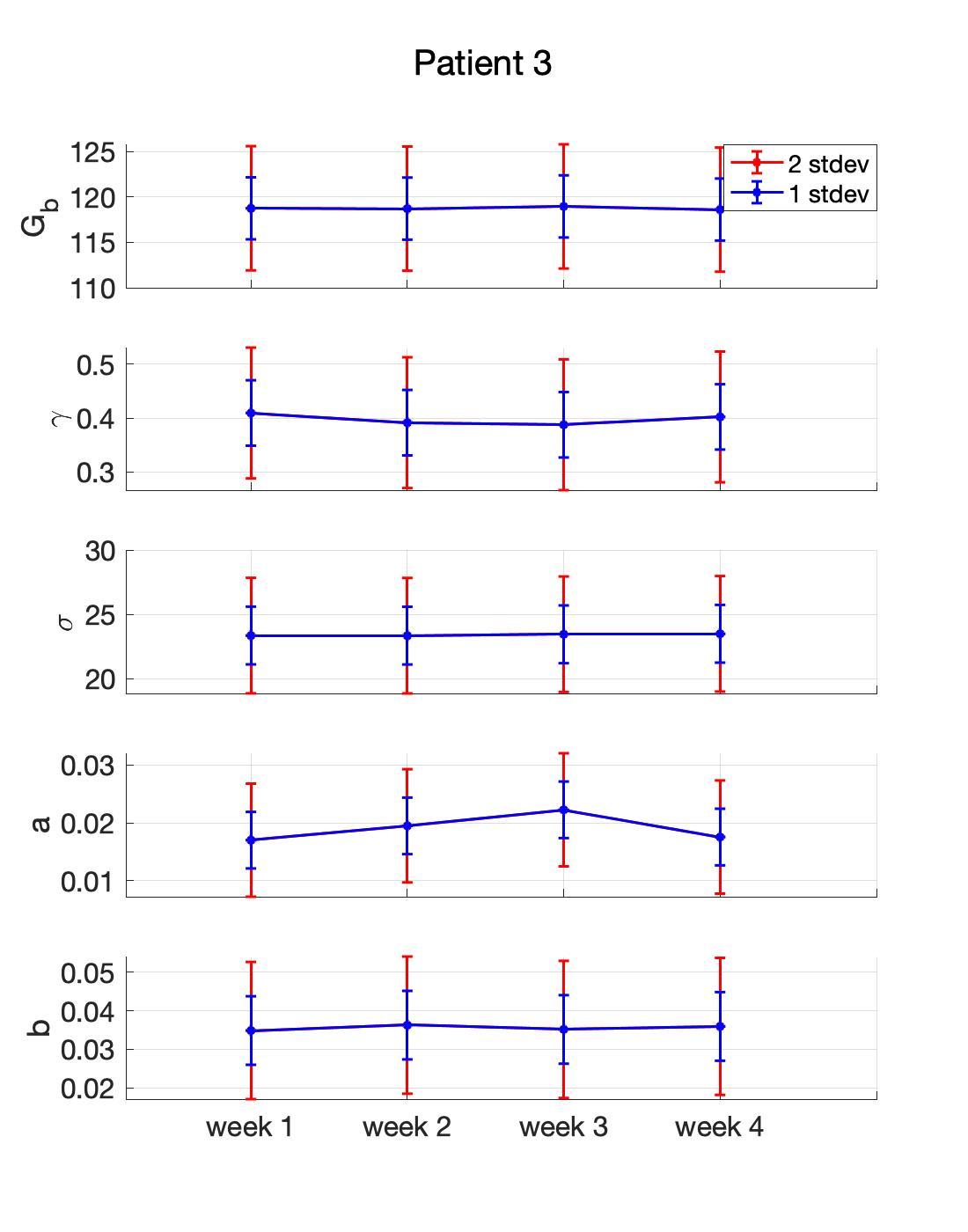}
		\caption{MCMC} \label{fig3F}
	\end{subfigure}
	\caption{Parameter estimation and uncertainty quantification in the T2DM setting. The top panel is obtained with optimization and the bottom panel is obtained with MCMC, both are in a patient-specific manner. It shows that the point estimates obtained with two approaches are very close to each other in most cases. Also the width of the 1- and 2-stdev intervals, which are obtained with Laplace approximation (optimization) and directly from the approximate posterior samples (MCMC), are also in agreement with each other. The parameter estimates and agree with real physiological values and the non-stationary behavior of the glucose dynamics of T2DM patients is reflected in the time-evolving behavior of the estimated parameters. All these features enforce the reliability of the parameter estimation results.}
	\label{fig3}
\end{figure}

To demonstrate that the estimated parameters are physiologically valid, consider Fig~\ref{fig3} where we see the point estimates and UQ intervals for all parameters and all three patients obtained with optimization and MCMC methods. The estimated basal glucose, $G_b$, values are in the ranges of $\sim95-105$ mg/dl, $\sim 105-140$ mg/dl, and $\sim 105-125$ mg/dl over the course of four weeks for patients 1, 2, and 3, respectively. These values are indeed in the expected ranges based on the BG measurements of these patients.

To show that the UQ intervals are potentially useful in practice, once again consider Fig~\ref{fig3}. The range of UQ intervals for each estimated parameter in most cases contains physiologically plausible parameter values that are tight enough to enforce the reliability of the point estimates. To quantify this statement we computed the coefficient of variation, defined as the standard deviation divided by the mean and can be interpreted as a measure of variability of the point estimator in this context. For $G_b$ and $\sigma$, which are the most influential parameters in characterizing the mean and variance of the model output, the coefficient of variation is in the $\sim 1-5\%$ band and $\sim 4-14\%$ band, respectively for all three patients. These results support the reliability of the point estimates that are used to form patient-specific models to describe dynamics of each patient.

We can see the robustness of the estimated parameter values by comparing parameter estimates using two different methods, optimization and MCMC. The results are shown in Fig~\ref{fig3}; the upper and lower panels show parameter estimates using optimization and MCMC, respectively. The point estimates as well as the corresponding UQ intervals for $G_b$ and $\sigma$ obtained with optimization and MCMC are very close to each other in most cases. Some parameters have more variation between methods; specifically, $\gamma$, $a$, and $b$ do show variation between the results obtained with optimization and MCMC methods. This variation does not seem to have substantial effect on the model's ability to represent patient dynamics. The overall result is a model whose ability to represent the data is relatively insensitive to parameter estimation techniques.

\subsubsection{Forecasting}\label{T2DM:forc}

We evaluate forecasting ability of the model in this setting along two pathways,
a face validity pathway that is mostly motivated by potential clinical decision-making,
and a more statistical-based pathway that is motivated by our desire to be quantitative.
In a sense, both evaluations address whether the data could plausibly be generated
by the model.

\begin{figure}[!ht]
	\begin{subfigure}{\textwidth}
	\centering
	\includegraphics[scale=0.18]{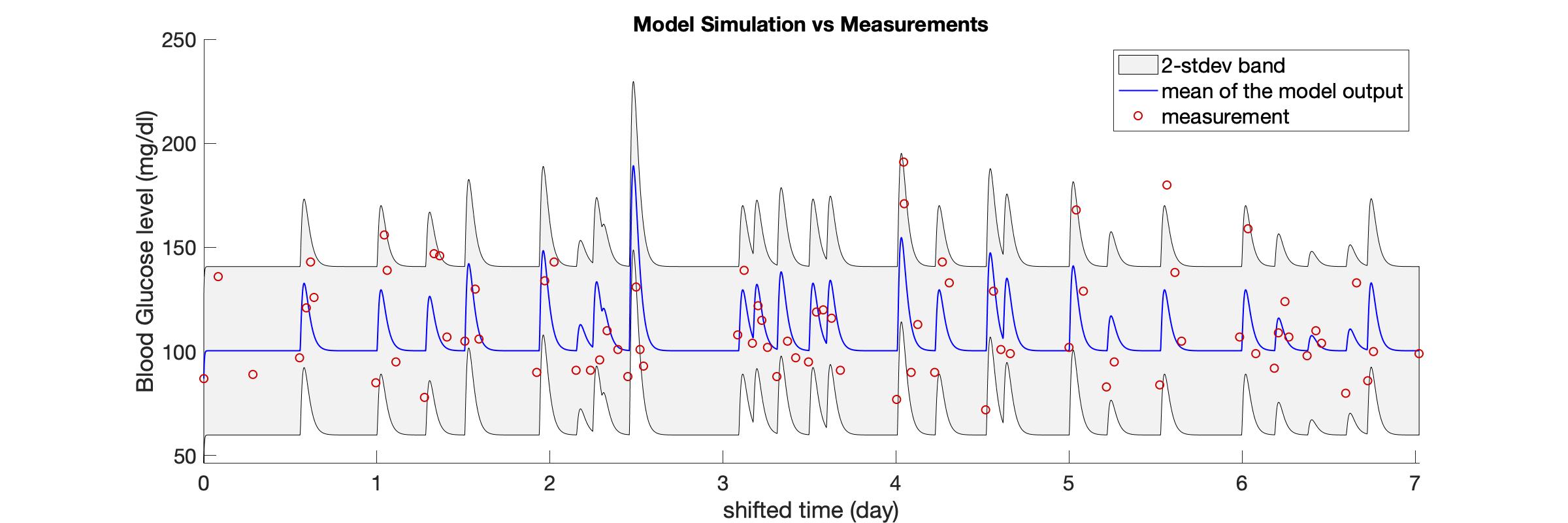}
	\caption{Model output with BG measurements}\label{fig4A}
\end{subfigure}\hspace{0cm}

\medskip

\begin{subfigure}{0.33\textwidth}
	\includegraphics[width =\linewidth]{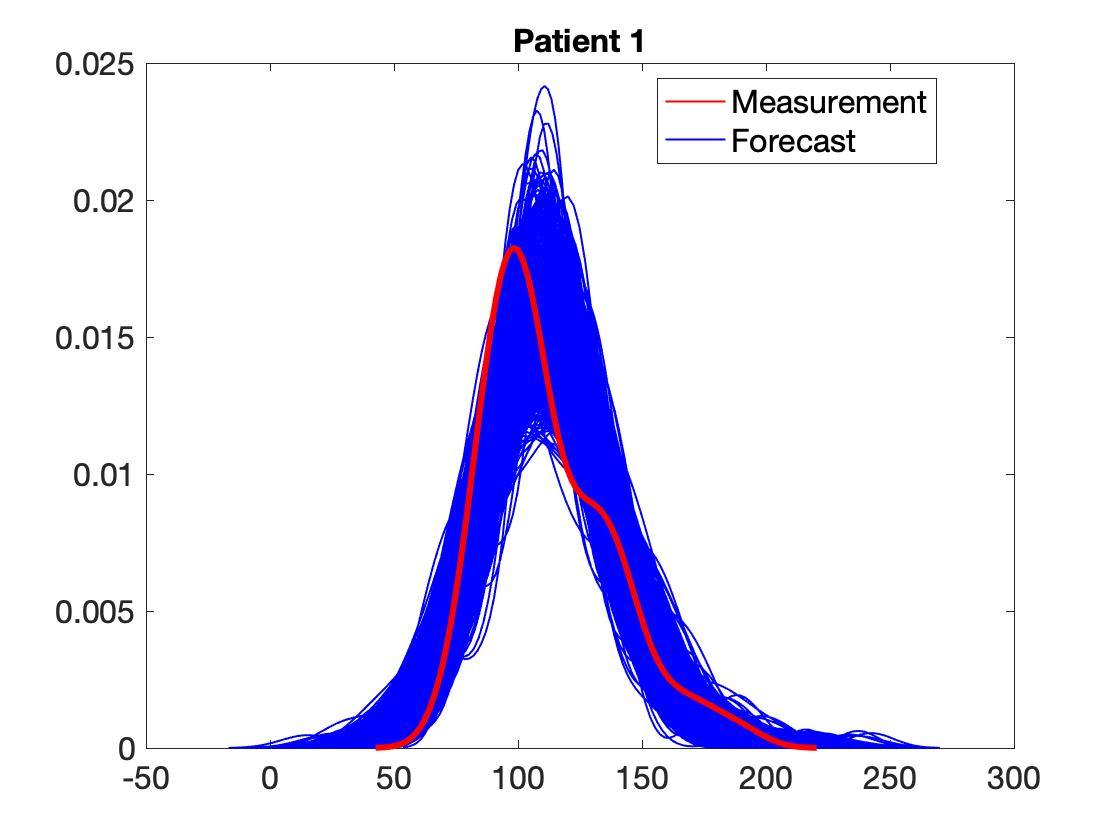}
	\caption{KDE - Patient 1}\label{fig4B}
\end{subfigure}\hspace{0cm}
\begin{subfigure}{0.33\textwidth}
	\includegraphics[width =\linewidth]{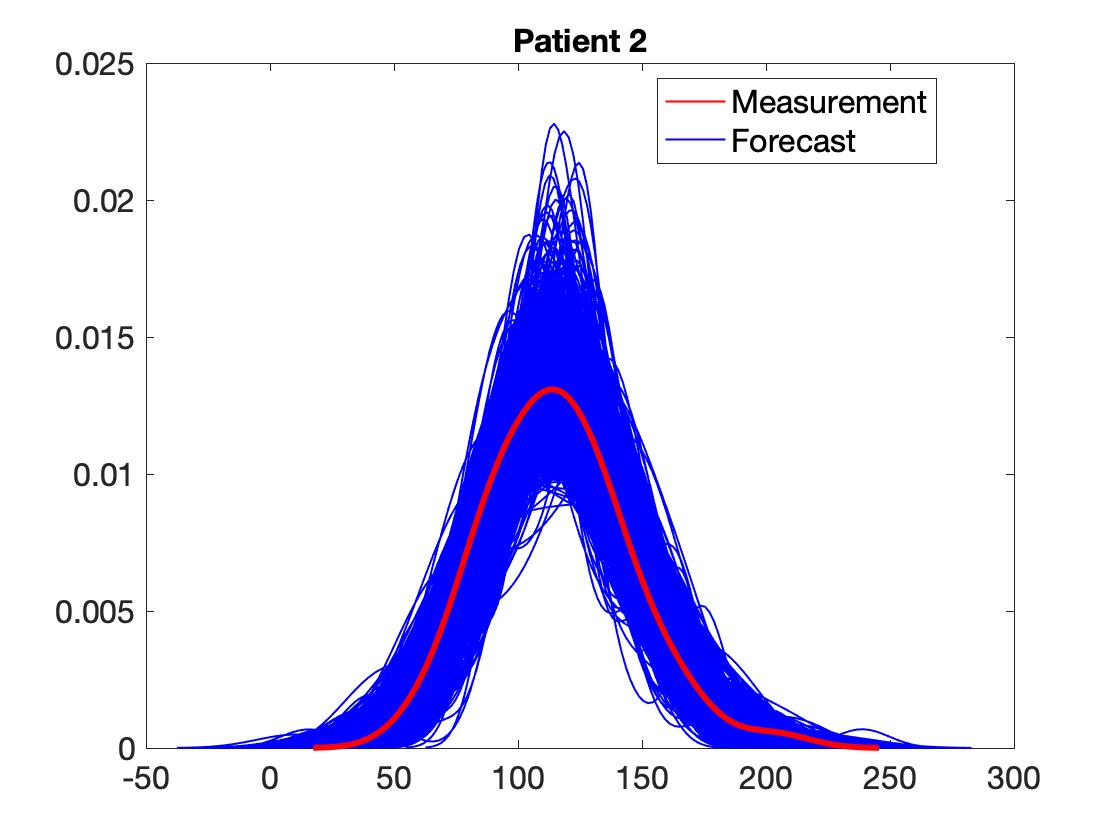}
	\caption{KDE - Patient 2}\label{fig4C}
\end{subfigure}
\begin{subfigure}{0.33\textwidth}
	\includegraphics[width =\linewidth]{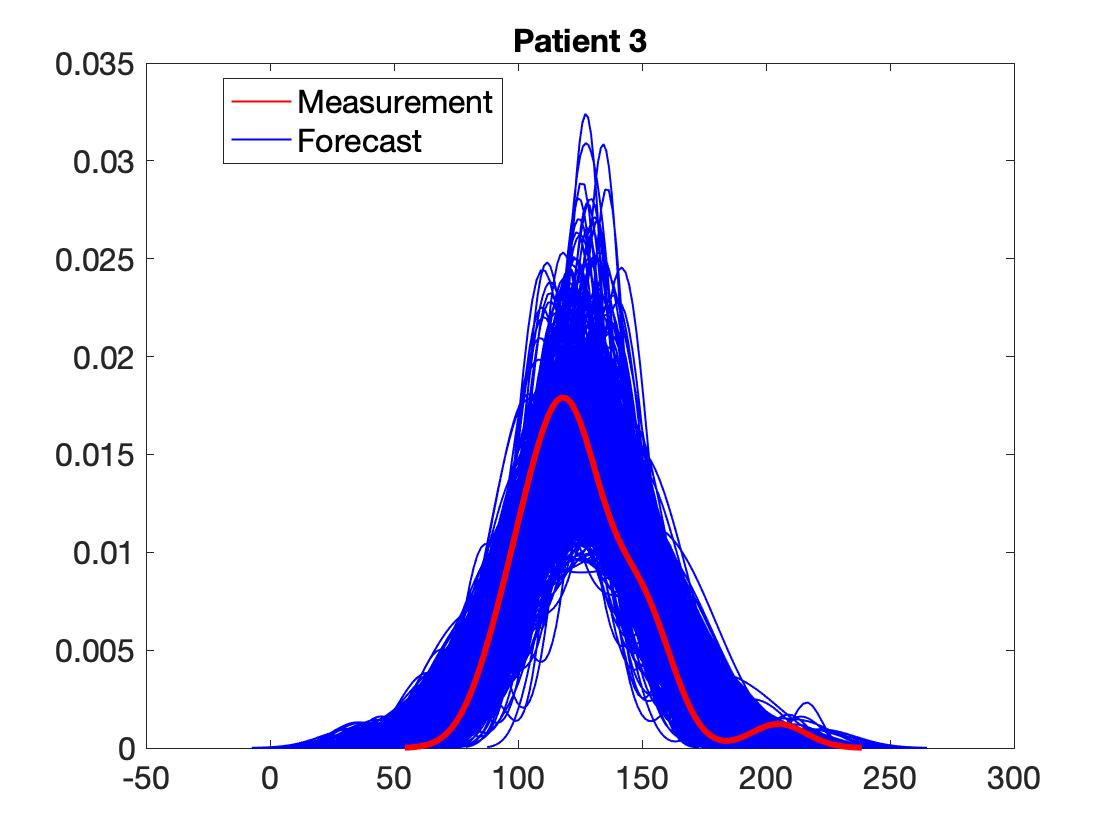}
	\caption{KDE - Patient 3}\label{fig4D}
\end{subfigure}
	\caption{On the top row, the model output of the estimated stochastic process is shown over the week of the training data along with the true BG measurements, and on the bottom row, kernel density estimate (KDE)  of 1,000 different realizations of the estimated model output and BG measurements are shown for each patient. In \textbf{(a)}, the red circles show true BG measurements, the blue curve shows the mean of the model output, and the gray area represents the estimated 2-stdev band around the mean. Comparison of the true BG measurements, which are assumed to be a realization of the model output, with the mean and 2-stdev band of the stochastic process---being the model output---along with the KDE plots in \textbf{(b)-(d)} implies that BG measurements could indeed be considered as a realizations of the random process.}
	\label{fig4}
\end{figure}

\begin{figure}[!ht]
	\begin{subfigure}{\textwidth}
		\centering
		\includegraphics[width=\linewidth]{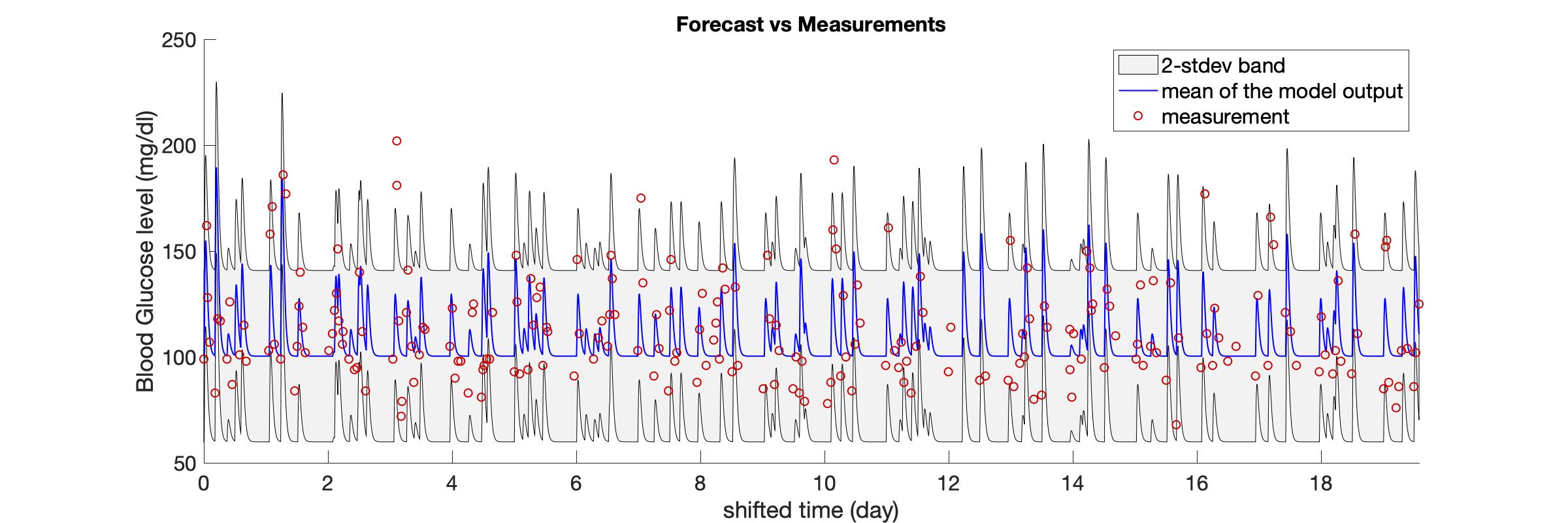}
		\caption{Patient 1} \label{fig5A}
	\end{subfigure}
	
	\medskip
	\begin{subfigure}{\textwidth}
		\centering
		\includegraphics[width=\linewidth]{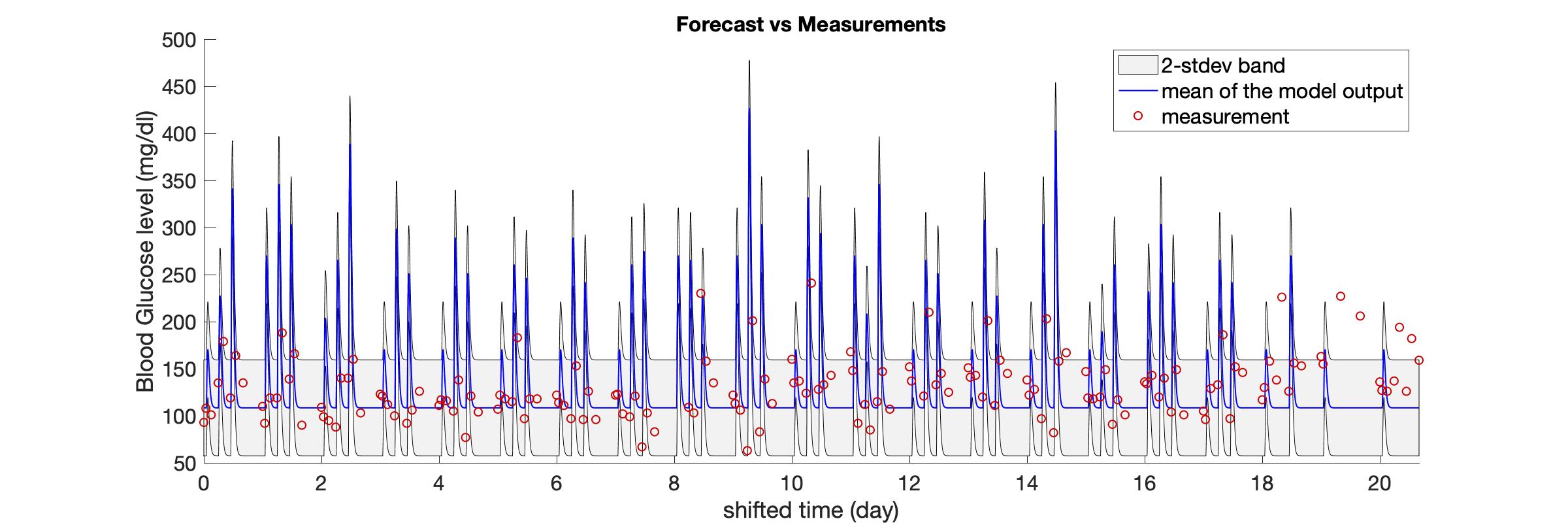}
		\caption{Patient 2} \label{fig5B}
	\end{subfigure}
	
	\medskip
	\begin{subfigure}{\textwidth}
		\centering
		\includegraphics[width=\linewidth]{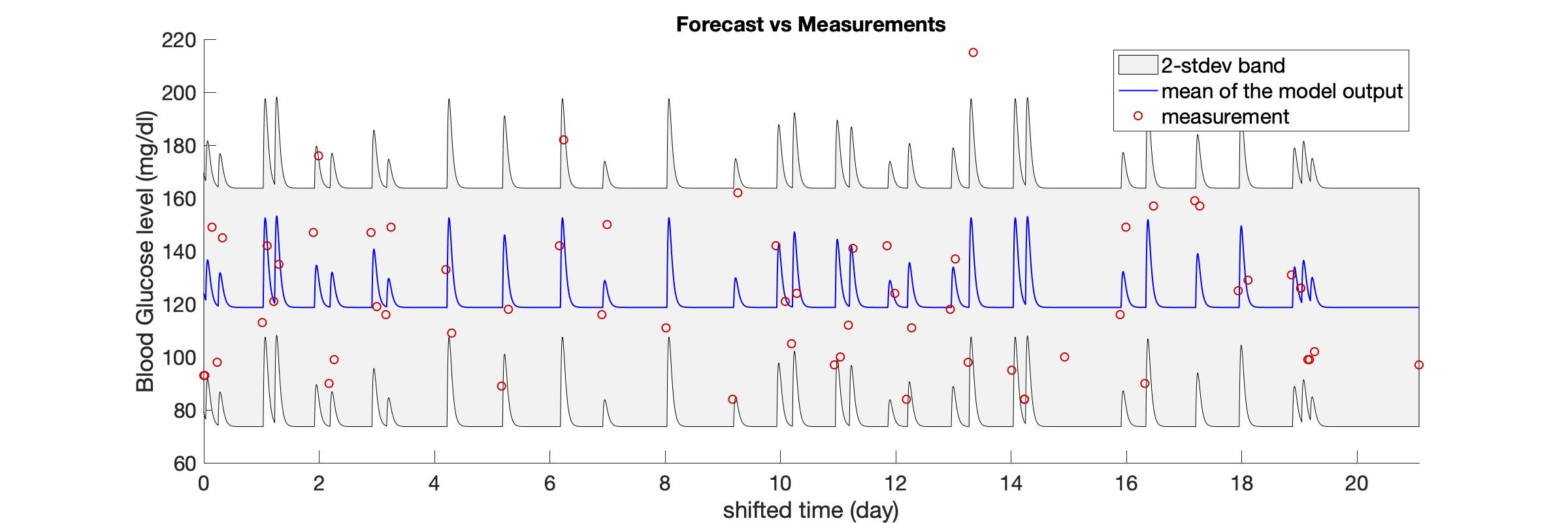}
		\caption{Patient 3} \label{fig5C}
	\end{subfigure}
	\caption{Forecasting results in the T2DM setting obtained via models formed by using the estimated parameters with the optimization approach. In each plot, the red circles show the true BG measurements, the blue curve shows the mean of the model output, and the gray region is the estimated 2-stdev band around the mean of the model output. These forecasting results show that the proposed model mean, when equipped with confidence bands found from standard deviations, estimate  the BG levels accurately, and in a patient specific way. This reinforces the claim that the model parameters could be used to provide information about the health condition of individual patients.}
	\label{fig5}
\end{figure}

The \emph{first} evaluation---face validity---is to consider whether the model can capture the dynamics qualitatively. Because the model's forecast is in the form of a distribution, the forecast we have to evaluate is anchored to the mean and standard deviation. In Figs~\ref{fig4A} and \ref{fig5}, the red circles are the BG measurements and from our modeling perspective are also a realization of the stochastic process whose mean is shown by the blue curve and variance is represented by the gray region. Fig~\ref{fig4A} shows the training time window for one of the patients and Figs~\ref{fig5A}-\ref{fig5C} show the test time window for each patient. An initial inspection of these figures implies that the model output seems to represent the data well. One important and challenging task of a forecast is accurate estimation of uncertainty.  Because of the idiosyncrasy of our stochastic model, uncertainty is quantified naturally using the covariance function of the model process. Fig~\ref{fig5} demonstrates the model's effectiveness in capturing relevant forecast uncertainty with two standard deviation (2-stdev) bands around the model mean; these bands capture most of the future BG measurement. These results are further quantified in Table \ref{tab3} that shows summary statistics for how often the future measurements were captured by the 2-stdev bands. Being able to contain $\sim 89-97\%$ of the true BG measurements in these confidence regions for all three patients is an indicator of this model’s predictive capability. Because of more dangerous consequences of hypoglycemia, we also check the percentage of measurements that are smaller than the lower 2-stdev band, i.e., missed by the 2-stdev band on the lower-end, over the test time window, which are 1.79\% (four measurements out of 224 total BG measurements), 0\%, and 0\% for patient 1, 2, and 3, respectively. These four measurements for patient 1 are 88, 94, 102, and 118 mg/dl, and the lower bound of the 2-stdev band for these measurements are estimated to be 94, 99, 105, and 123 mg/dl. Also, this patient had total of 31 BG measurements in the range of 68-88 mg/d, and the estimated 2-stdev band missed only one of BG measurements (88 mg/dl) in that range and estimated the possibility of occurrence of all the remaining ones. This result shows that model could provide decision support for the possible occurrence of hypoglycemia. Thus, this model is providing substantial forecasting information beyond what is available given the data alone.

The \emph{second} evaluation quantifies how plausible it is that the data we observe
could have originated from the model. We quantify this plausibility using the
two-sample Kolmogorov-Smirnov (KS) test. To start, Fig~\ref{fig4B}-\ref{fig4D}
show the kernel density estimates (KDEs) obtained from the BG measurements (red curve)
and from 1,000 independent realizations of the estimated stochastic process (blue curves)
over the training time window for each patient. The KDEs in Fig~\ref{fig4B}-\ref{fig4D}
support the idea that the BG measurements could be assumed to be drawn from the
distribution given by the estimated model output. To perform the two-sample
Kolmogorov-Smirnov (KS) test we created  datasets by resampling 10,000 independent
realizations of the model output at the BG measurement times and performed the test
using each generated sample against BG measurements with the {\ttfamily kstest2} function in
{\ttfamily MATLAB} with 1\% significance level. We performed this procedure over one-week of
training window and three-week of test window for each patient separately. Note that the null
hypothesis states that the two samples are drawn from the same distribution and
\textit{not rejecting the null hypothesis} supports that our model could accurately represent the
distribution of the measurements. Moreover, the null hypothesis here is a distributional one so
that re-ordering measurements or forecasts will have no effect on the KS test.

Out of 10,000 different samples in each case, the rejection rates were $0.55\%$, $0.02\%$,
and $0.59\%$ over the training time window and they were $1.65\%$, $67.54\%$, and $0.38\%$
over the test time window, respectively for patients 1, 2, and 3. First, observe that the rejection
rates are much smaller over the training time window. This is expected as the random samples
used against the BG measurements in the KS test are generated by the model output estimated
using the same BG measurements. However, the model output used to generate samples over the test time window was obtained only using the patient-specific model, which is trained by the training data and nutrition intake data of those patients over the test time window. Therefore, even though we have a high rejection rate for patient 2 over the test time window, those much smaller numbers for patients 1 and 3 are still reassuring and show that our initial assumption, which is that our simplified stochastic model can describe the BG values, is a valid assumption in this setting.

While the KS test establishes the distributional similarity between the data and our fitted model, we also evaluate pointwise correlations to establish the validity of the model's predicted
dynamics; i.e., responses to meals. We report the Pearson correlation coefficients in
Table \ref{tab3}, and show substantial positive correlation. This indicates that our model
is superior to a constant statistical model of the data distribution.

Finally, we see from Fig~\ref{fig5B} that the mean of the model output exhibits unusually high peak BG values after the meals. In addition, the Kolmogorov-Smirnov test has a high rejection rate for this patient. Investigation of parameter values shown in Fig~\ref{fig3A}-\ref{fig3C} reveals that there is an order of magnitude difference between estimated gamma values for patient 2 and for patients 1 and 3. Since $\gamma$ represents the decay rate to the patient's basal glucose value, we hypothesized the reason for not estimating the gamma parameter accurately for this patient could be related to their BG measurement pattern. To investigate, we checked the time difference between the recorded meal times and the first BG measurement times after each meal. We found that patient 2 had 18 meals over their training time window and that time difference was exactly 2 hours for each meal. Patients 1 and 3 had variability among their measurement times. We believe such a regular measurement pattern without any variability is the reason for not being able to estimate the gamma parameter, representing the decay rate to the basal glucose value. In addition, we believe this is also the reason for the high rejection rate for Kolmogorov-Smirnov test for patient 2. We provide more detail about the BG measurement pattern of these patients in the Supplementary Material.

\subsubsection{Comparison of Forecasting Accuracy with Longitudinal Diabetes Pathogenesis Model}\label{t2dm_comparison}

In this section, we compare the forecasting accuracy of the T2DM version of the MSG
model with a well-known model, the longitudinal diabetes pathogenesis (LDP) model
developed by Ha \& Sherman \cite{ha2020type} and a simple mean-variance model.
The LDP model is developed to understand different pathways of T2DM pathogenesis. It represents the metabolic state of T2DM patients at any time during the disease progression over years. The model consists of four differential equations and 11 model parameters. We do not attempt
to estimate all these parameters as it is not feasible with the
available sparse data. We perform the same forecasting task by
estimating three different sets of parameters, $\{\sigma, SI\}, \{\sigma,SI,hepaSI\}, \{\sigma,SI,hepaSI,r20\}$, and setting the remaining parameters at known
default values.

\begin{table}[!ht]
	\centering
	\begin{tabular}{|c|l|c|c|c|c|c|}
		\hline
		\multicolumn{7}{|c|}{Patient 1}                                                                                                                      \\ \hline
		\multicolumn{2}{|l|}{}                                                  & 1-std \% & 2-std \% & RMSE & MPE & CORR \\ \hline
		\multicolumn{2}{|c|}{MSG Model}                             & 75.45             & 93.30             & 20.12         & 12.97		&	0.5680        \\ \hline
		\multirow{3}{*}{LDP Model} & $\sigma,$ SI              & 41.96             & 65.62             & 22.89         & 13.72		& 0.5090        \\ \cline{2-7}
		& $\sigma,$ SI, hepaSI      & 40.18             & 66.96             & 22.00         & 13.77		& 0.4995        \\ \cline{2-7}
		& $\sigma,$ SI, hepaSI, r20 & 43.30             & 65.62          & 22.09         & 13.59		& 0.5737       \\ \hline
		\multicolumn{2}{|c|}{Mean-Variance Model}		&	73.66		&	95.98		&	24.48		&	16.97		&	0		\\ \hline
		\multicolumn{7}{|c|}{Patient 2}                                                                                                                    \\ \hline
		\multicolumn{2}{|l|}{}                                                  & 1-std \% & 2-std \% & RMSE & MPE & CORR \\ \hline
		\multicolumn{2}{|c|}{MSG Model}                             & 63.29             & 89.24             & 33.52         & 17.35		& 0.3674        \\ \hline
		\multirow{3}{*}{LDP Model} & $\sigma,$ SI              & 20.89             & 37.34             & 39.54         & 21.00		& 0.3122       \\ \cline{2-7}
		& $\sigma,$ SI, hepaSI      & 15.82             & 32.91             & 44.18         & 24.20		& 0.3019        \\ \cline{2-7}
		&$\sigma,$ SI, hepaSI, r20 & 18.35             & 33.54          & 40.38         & 21.71		 & 0.3536        \\ \hline
		\multicolumn{2}{|c|}{Mean-Variance Model}		&	68.99		&	90.51		&	35.54		&	18.17		&	0		\\ \hline
		\multicolumn{7}{|c|}{Patient 3}                                                                                                                       \\ \hline
		\multicolumn{2}{|l|}{}                                                  & 1-std \% & 2-std \% & RMSE & MPE & CORR \\ \hline
		\multicolumn{2}{|c|}{MSG Model}                             & 51.61             & 96.77             & 24.27         & 17.12		&  0.4759        \\ \hline
		\multirow{3}{*}{LDP Model} & $\sigma,$ SI              & 29.03             & 50.00           & 32.20         & 18.98  &  0.3750        \\ \cline{2-7}
		& $\sigma,$ SI, hepaSI      & 30.65             & 53.23             & 32.88         & 18.69		&  0.4032       \\ \cline{2-7}
		& $\sigma$, SI, hepaSI, r20 & 19.35             & 46.77           & 33.43         & 19.81		&  0.4019        \\ \hline
		\multicolumn{2}{|c|}{Mean-Variance Model}		&	58.07		&	95.16		&	26.96		&	18.74		&	0		\\ \hline
	\end{tabular}
	\caption{Comparison of the forecasting results with three different models. For each different case of the LDP model the results in the corresponding row shows which parameters are estimated during the whole forecasting experiment. We obtain better forecasting accuracy with the MSG model than with the LDP model and mean-variance model, in general. Furthermore, for the LDP model, the forecasting accuracy decreases as the number of parameters being estimated increases.}
	\label{tab3}
\end{table}

The experiment in this setting will be the same as described above in Section \ref{T2DM:forc}. For a fair
comparison, mean-variance model corresponds simply to computing the sample mean and variance
from the training data and to using the mean as the point estimator over the test time
window and the variance for uncertainty quantification in the forecast. On the other hand,
the LDP model consists of a set of coupled ODEs. To estimate the unknown model
parameters and forecast BG levels with the LDP model we used the constrained
Ensemble Kalman Filter (EnKF) algorithm, \cite{albers2019ensemble}.
We coded the algorithm on {\ttfamily MATLAB} for parameter estimation
and BG forecasting using the constrained EnKF method based on the LDP model.
We used {\ttfamily MATLAB}'s ODE solver {\ttfamily ode23} to solve the LDP model numerically.

Note that we use the constrained EnKF algorithm because it is validated to provide accurate
forecasting results with complex ODE models \cite{albers2019ensemble}. Moreover, the \textit{ensembles} of state estimates could be used for uncertainty quantification.
However using a filtering algorithm requires exploiting all the data collected up until the
forecasting time point; unlike the optimization algorithm paired with MSG model, which
could use data collected only over the training time window to train the model and then
simulate over the test time window for forecasting. Note that with LDP model - EnKF
algorithm pair, we used all the data contained in the training and test time windows. Then,
we used the BG forecasting values over the test time window for comparison.
The comparison results are shown in Table \ref{tab3}.

The results in Table \ref{tab3} show that the MSG model provides better accuracy in
forecasting future BG levels in T2DM patients than all variants of
the LDP model and mean-variance model when compared holistically.

\emph{First}, MSG model achieves smaller RMSE and  MPE than
all different variations of the  LDP model and mean-variance model for all three
patients, demonstrating that mean of the MSG model output is also preferable as a point estimator.

\emph{Second}, we see the advantage  of using a stochastic model which
inherently quantifies the level of certainty in the BG predictions. It is worth noting
that the MSG model is based on learning parameters of a stochastic
model, whilst the LDP quantifies uncertainties by learning an ensemble
of parameters and states; this may contribute to the differences between them at
the level of uncertainty prediction. The percentages in Table \ref{tab3}
show that the MSG model is substantially better than the LDP model in
capturing the true measurements in the corresponding confidence bands
whereas it is not as good as mean-variance model for these percentages. Nevertheless,
the comparison of the correlation values over the test time window supports the better forecasting accuracy with the
MSG model. In summary the MSG model is preferable to the LDP and mean-variance models
for decision making within the context we use here, as it possesses the good
features when compared with different type of models. When compared
with a mechanistic model as itself (the LDP model here), the MSG model gives
a better point forecasts and better confidence bands, enabling knowledge
of possible high and low values for future BG levels. On the other hand,
when compared with the most simple data driven model, it still provides
better overall forecasting accuracy.

\subsection{ICU}\label{num_res_ICU}

We now move to the more complex and difficult case of modeling and forecasting glycemic dynamics in the ICU, where non-stationarity is manifest on much shorter time-scales. Parameter estimation and forecasting are, in general, harder in the ICU context because of the characteristics of ICU patients as explained in Section \ref{clinical_setting}. A detailed explanation about how the MSG model represent the dynamics in ICU setting is provided in the Supplementary Material.

\subsubsection{Parameter Estimation}\label{param_est_icu}

\begin{figure}[!ht]
		\begin{subfigure}{\textwidth}
		\centering
		\includegraphics[scale=0.12]{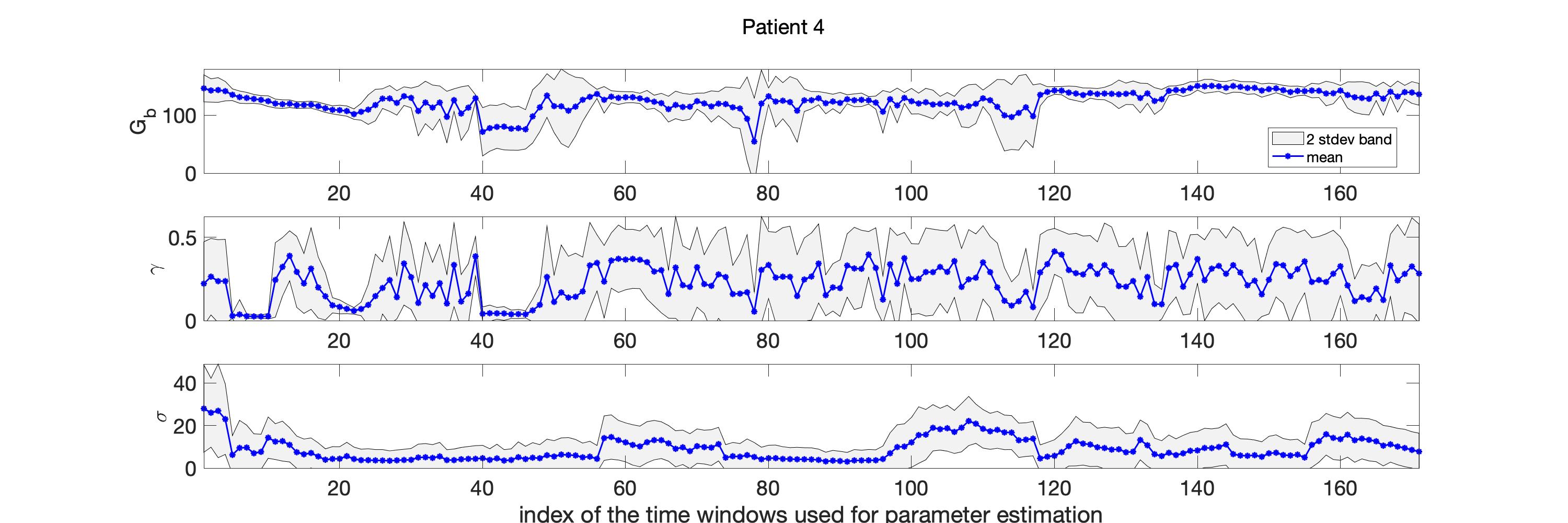}
		\caption{patient 4} \label{fig6A}
	\end{subfigure}
	
	\medskip
	%	\hspace{-1cm}
	\begin{subfigure}{\textwidth}
		\centering
		\includegraphics[scale=0.12]{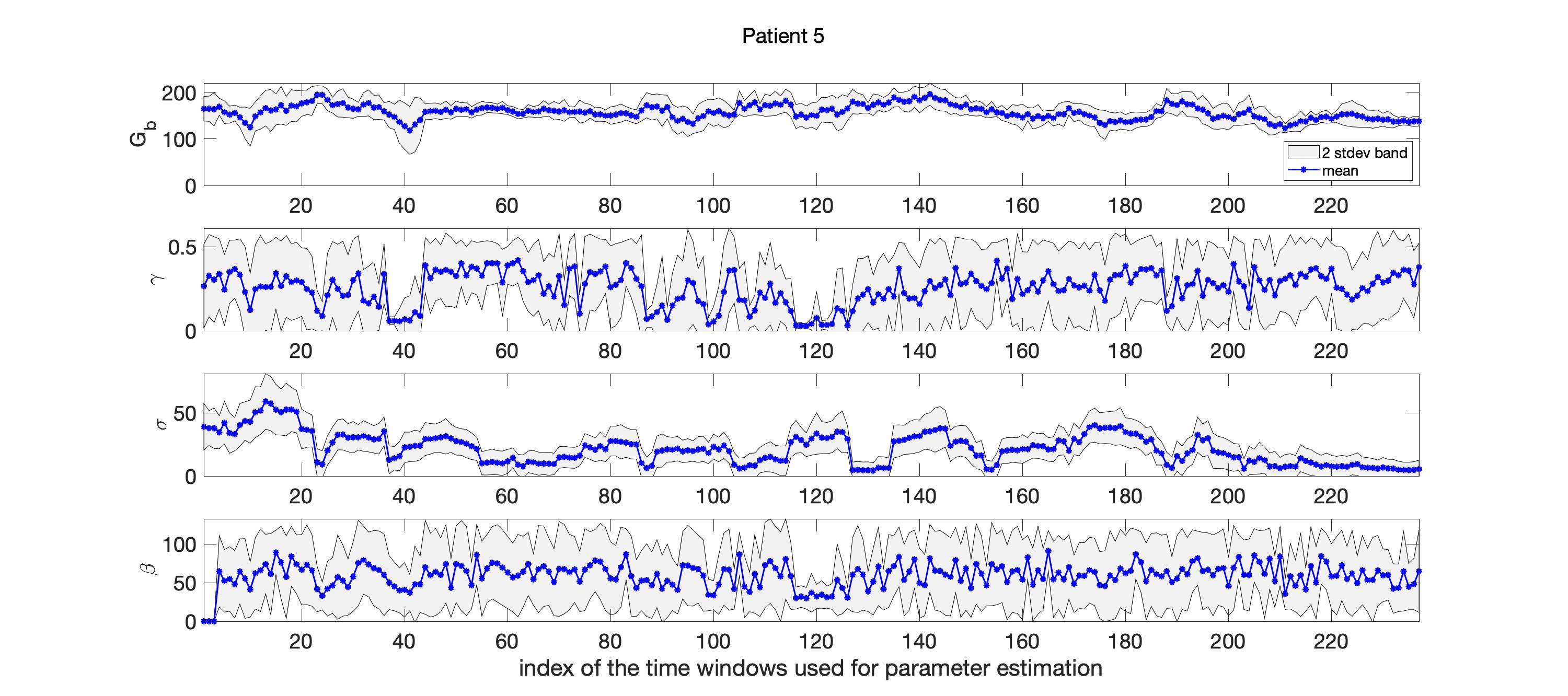}
		\caption{patient 5} \label{fig6B}
	\end{subfigure}
	
	\medskip
	%	\hspace{-1cm}
	\begin{subfigure}{\textwidth}
		\centering
		\includegraphics[scale=0.12]{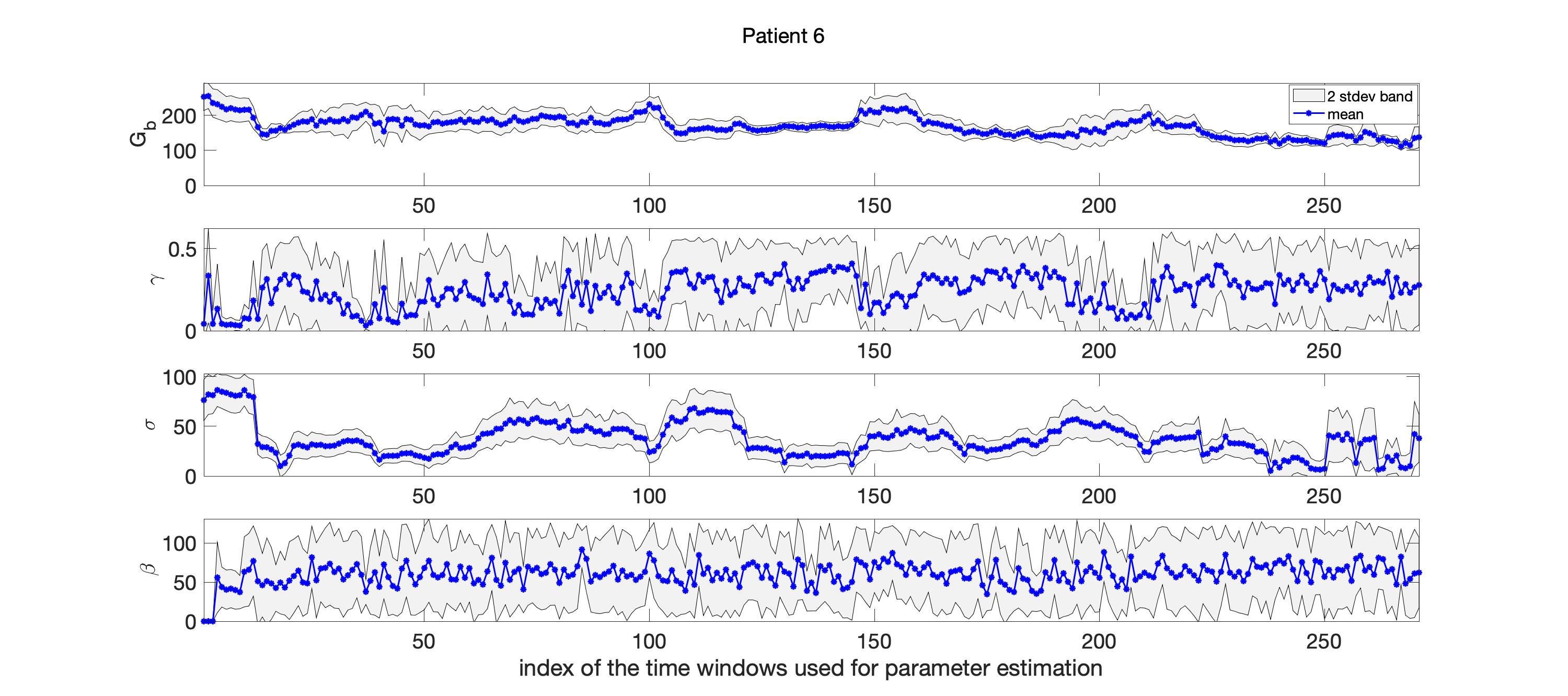}
		\caption{patient 6} \label{fig6C}
	\end{subfigure}
	\caption{Parameter estimation and uncertainty quantification results in the ICU setting obtained with MCMC. In each plot, the blue stars are the point-estimate of each parameter and the gray area is the 2-stdev band around the point-estimates (both obtained from the resulting random samples). The estimated model parameters exhibit biophysically realistic values. Also, 1- and 2-stdev bands enforces the reliability of the estimated parameters, especially, $G_b$ and $\sigma$, which are the most influential parameters in predicting the mean and variance of BG levels.}
	\label{fig6}
\end{figure}

The difficulties presented in the ICU setting are reflected in our parameter estimation results. Despite these complexities, our results exhibit four substantive pieces of evidence which support the validity of the model and its potential effectiveness for understanding the physiological state of ICU patients and for forecasting. \emph{First}, the model captures the dynamics reflected in the parameter estimates with sparse data. \emph{Second}, the estimated model parameters, which have the most influence in resolving the mean and variance of the BG level, are physiologically valid in most of the cases. \emph{Third}, the changes in the parameter estimation results over moving time windows are realistic and reflective of the expected non-stationary behavior of ICU patients. And \emph{fourth}, the UQ results show that the parameters (basal glucose rate, $G_b$ and the model standard deviation, $\sigma$), which have the most influence in resolving mean and variance of BG levels are estimated with more certainty. Having tighter bands around the point estimates for these parameters indicates the robustness of the estimation. We explain these claims in detail in the following paragraphs. Fig~\ref{fig6} contains evidence supporting all these results and Table \ref{tab2} shows the sparsity of the BG measurements.

With the complexity of ICU data in mind, consider the parameter estimation results. Figure \ref{fig6} shows the parameter estimates over moving time windows with length of 24 hr for each ICU patient obtained with MCMC approach. The mean of each chain is shown using blue stars. These parameter estimates are physiologically plausible for all three patients except in a small number of cases. For example, estimates of the basal glucose rate, $G_b$, were around $\sim 105-145 \text{mg/dl}$, $\sim 140-180 \text{mg/dl}$, $\sim 135-210 \text{mg/dl}$, for patients 4, 5, and 6, respectively, all plausible values given the patient's data. As shown in the Supplementary Material, it was not possible to compute good estimates for parameters $\gamma$ and $\beta$ in some of the cases.

Fig~\ref{fig6} also shows that the time evolution of the estimated parameters is realistic within the ICU context. In ICU the training time windows move in positive (increasing time) direction of measurements---given a measurement the model is estimated using the previous 24 hours of data, $\sim 14$ data points to forecast the future measurement whenever it comes---so that the consecutive time windows have an overlap of 20-23 hours. This means that the model varies relatively continuously between consecutive time windows. This relative continuity is reflected in Fig~\ref{fig6} that shows the time evolution of estimated parameters for all three patients. Even though the health condition of the ICU patients can change rapidly, the estimated parameters do not change wildly (in most of the cases), reflecting the expectation under these settings. Nevertheless, the patients are clearly non-stationary and the observed evolution of the parameter estimates, shown in Fig~\ref{fig6}, reflects this non-stationarity.

And finally, as was the case in the T2DM setting, the model is relatively robust to the methods used to estimate it; however, as can be inferred from the discussion above about parameters and their face validity to physiology, the ICU formulation of the model can have more complex parameter estimation issues compared to the T2DM setting. In particular, in the ICU setting there are some cases where the Laplace approximation does not work well because the parameter misfit solution surface is flat in some parameter directions -- a reflection of identifiability issues. We provide more insight about this issue in Supplementary Material. Even though the point estimates for each patient and parameter pair by MCMC and optimization are close to each other, since UQ results are more meaningful by MCMC, we present the plots obtained by MCMC. In general we observe that $G_b$ and $\sigma$, both allow for more robust estimation compared to the estimation of $\gamma$ and $\beta$. The robustness of the estimation of $G_b$ and $\sigma$ is important for clinical applications because the mean and variance are what is used for glycemic management. As a demonstration of the robustness of $G_b$ and $\sigma$, consider Fig~\ref{fig6}. Here we can see the 2-stdev band around the mean for $G_b$ and $\sigma$ is tighter than the 2-stdev bands for $\gamma$ and $\beta$ for all three patients. Remember that both $\gamma$ and $\beta$ are related to the glucose removal rate from the blood. This is, perhaps, an indicator of an identifiability issue for these parameters. But it is also true that we are indeed less certain about this physiology; glucose can be removed at different rates by different physiological processes, e.g., liver versus adipose tissue, and we are not resolving these physiological subsystems. Moreover, due to the non-stationary and sparse nature of the data in the ICU setting, it is harder to estimate some of the model parameters accurately. Separating these inference issues is not possible given the data presently collected in these settings. Nevertheless, the parameters that play a key role in resolving the mean and variance of the BG dynamics can be estimated accurately up to the desired level.

\subsubsection{Forecasting}

Forecasting results in the ICU setting are indicative of two major features of this model: (i) we can capture the trend of BG measurements through the mean of the model and (ii) we can estimate the variance of the BG measurements accurately. Once again, since resolving mean and variance of BG dynamics is central to glycemic management, these results show potential usefulness of this model in the ICU context.

\begin{figure}[!ht]
	\begin{subfigure}{\textwidth}
		\centering
		\includegraphics[scale=0.15]{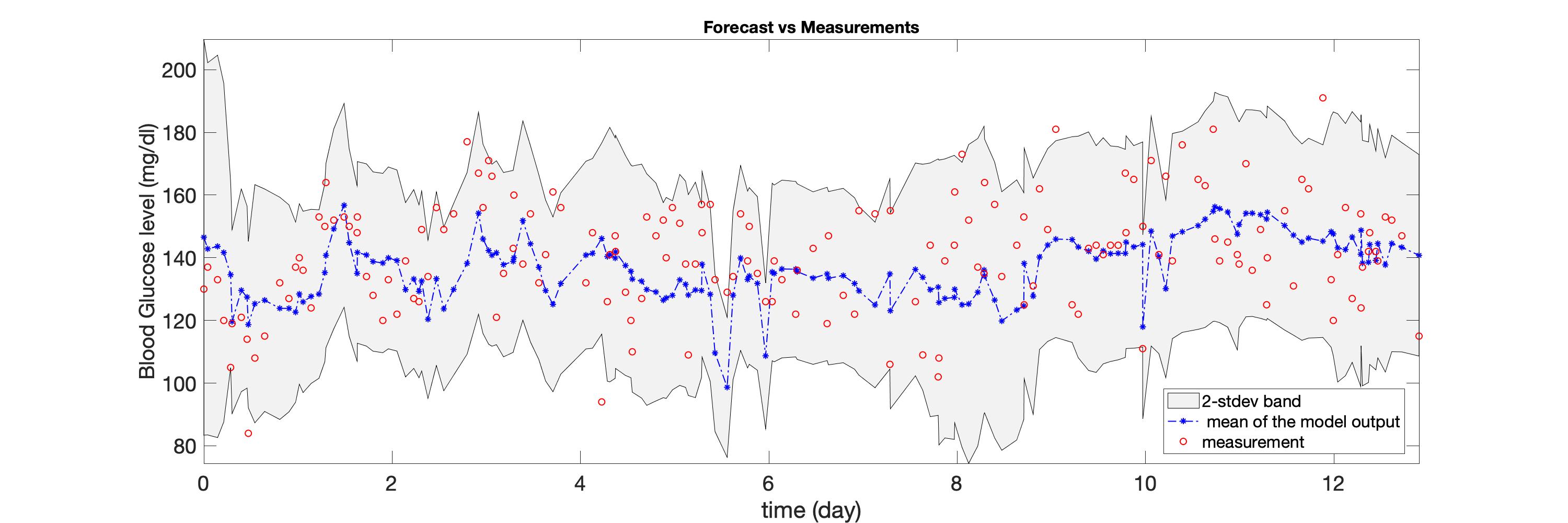}
		\caption{patient 4} \label{fig7A}
	\end{subfigure}

	\medskip
	%	\hspace{-1cm}
	\begin{subfigure}{\textwidth}
		\centering
		\includegraphics[scale=0.15]{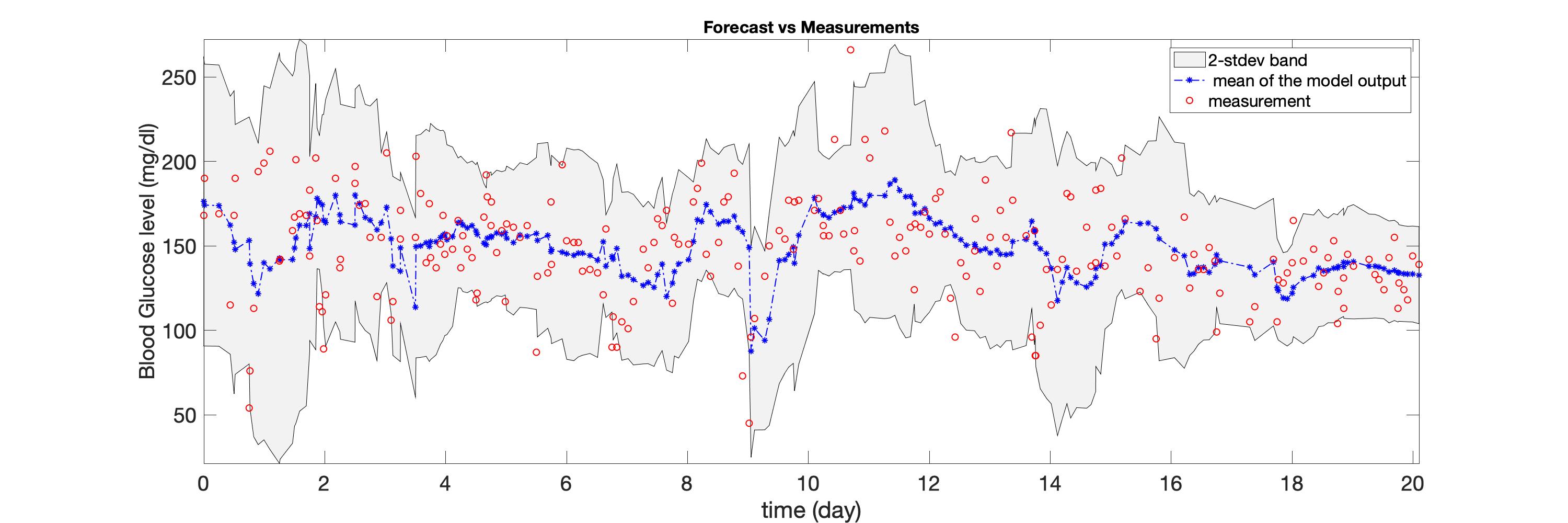}
		\caption{patient 5} \label{fig7B}
	\end{subfigure}

	\medskip
	%	\hspace{-1cm}
	\begin{subfigure}{\textwidth}
		\centering
		\includegraphics[scale=0.15]{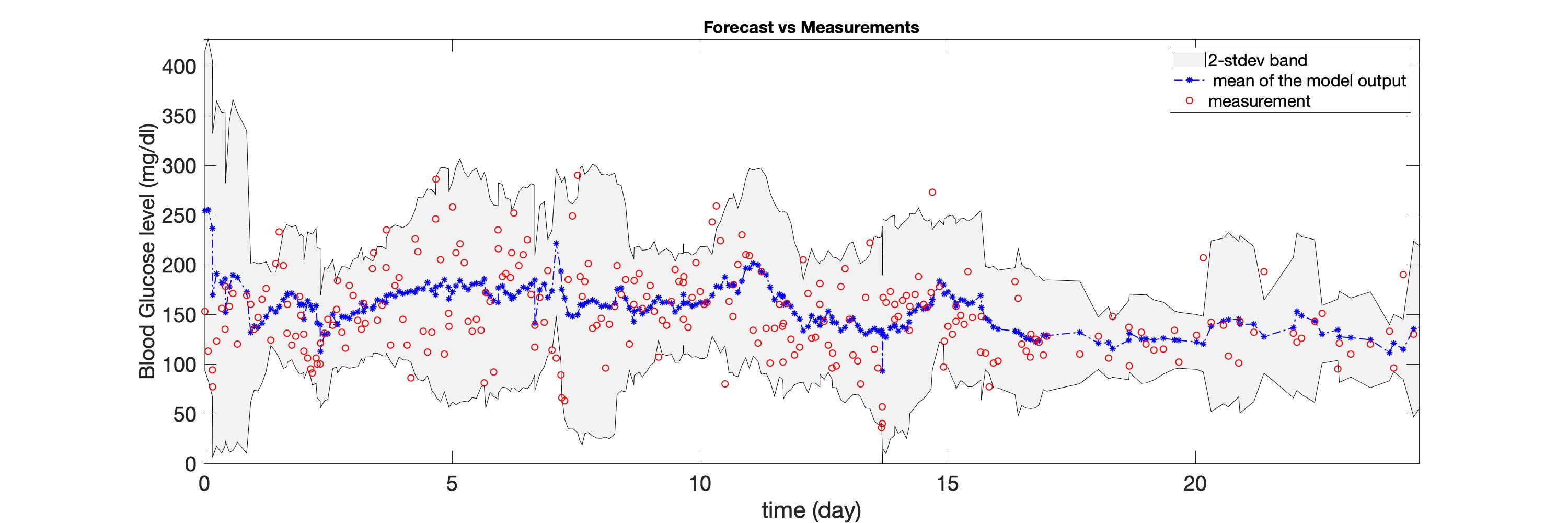}
		\caption{patient 6} \label{fig7C}
	\end{subfigure}
	\caption{Forecasting results obtained based on parameters estimated with MCMC, in the ICU setting. In each plot, the red circles show the true BG measurements, the blue stars show the mean of the model output, and the gray region shows the 2-stdev band around this mean. These results are, in general, very close to those obtained using the optimization approach, and the most relevant properties are shared by them both. Obtaining similar results with another numerical solution technique based on the same mechanistic model shows the reliability of the model and estimated model parameters.}
	\label{fig7}
\end{figure}

Fig~\ref{fig7} demonstrates that the forecasted mean of the model output encapsulates the essence of the behavior of BG measurements for all three patients. In each of the plots in Fig~\ref{fig7}, the red circles show the BG measurements, the blue stars are the mean of the model, and the gray region is the 2-stdev band around the mean, each one obtained separately with the corresponding patient-specific model.

To observe the effectiveness of this model in estimating the variance of the BG measurements accurately, consider Fig~\ref{fig7} and Table \ref{tab4}. Fig~\ref{fig7} shows the ability of the model to estimate the variance in glycemic dynamics visually where a large number of true BG measurements are contained in the gray regions that represent the \emph{forecasted} 2-stdev bands around the forecasted mean. These results are quantified in Table \ref{tab4} which contains summary statistics both for optimization and MCMC methods and demonstrate the forecasting accuracy of the MSG model and imply potential use in the ICU for glycemic management. Note that given the forecast, data, and the fundamental challenges of the ICU setting, we should expect forecast UQ bands to be quite large. But, this does not mean it is not valuable to clinicians but rather means that we can provide a realistic estimate of the uncertainty in an extremely volatile and poorly measured system.

\subsubsection{Comparison of Forecasting Accuracy With The ICU Minimal Model}\label{icu_comparison}

In this section, we use the ICUMM and mean-variance model in a similar
manner as in T2DM context for the comparison of the forecasting results.
The ICUMM is a physiological model that represents the glucose-insulin
dynamics of ICU patients and was developed to be used for glycemic
management in ICU. The model consists of four coupled differential equations
and has 12 model parameters. One of those
model parameters is used for the purpose of having units equal on
both sides of the equation and set to 1. Two of the model parameters
represent the volume of glucose and insulin distribution space and
are set to nominal values from the literature. This leaves us with nine
unknown model parameters to be estimated.
For parameter estimation and BG forecasting, we used 
the constrained EnKF method, which enables us to obtain confidence
bands for the forecasting results using the ensembles. We implemented
the BG forecasting algorithm using the EnKF method based on the ICUMM on
{\ttfamily MATLAB}. We used {\ttfamily MATLAB}'s ODE solver {\ttfamily ode45} to obtain
the solution of ICUMM. The mean-variance model simply uses the mean
and variance of BG measurements over the training time window for forecasting.

\begin{table}[!ht]
	\centering
	\begin{tabular}{|c|c|c|c|c|c|c|}
		\hline
		\multicolumn{7}{|c|}{Patient 4}                             \\ \hline
		\multicolumn{2}{|l|}{} & 1-std \% & 2-std \% & RMSE &  MPE	& CORR	 \\ \hline
		\multirow{2}{*}{MSG Model}	& optimization  & 56.14		& 89.47			& 17.27       & 10.29         & 0.3177        \\ \cline{2-7}
		& MCMC	&	60.82	& 91.81		& 18.60		& 11.07		& 0.2341	\\ \hline
		\multicolumn{2}{|c|}{ICUMM}		& 45.61			& 80.12         & 17.35       & 9.87         & 0.3232        \\ \hline
		\multicolumn{2}{|c|}{Mean-Variance Model}		& 59.06		& 94.15		& 18.07		& 10.96		& 0.1753	\\ \hline
		\multicolumn{7}{|c|}{Patient 5}                             \\ \hline
		\multicolumn{2}{|l|}{} &1-std \% & 2-std \% & RMSE & MPE & CORR \\ \hline
		\multirow{2}{*}{MSG Model}	& optimization  & 58.65		& 83.97			& 33.27      & 18.91         & 0.2751        \\ \cline{2-7}
		& MCMC	& 64.14		& 88.19		& 30.95		& 18.51		& 0.2773	\\ \hline
		\multicolumn{2}{|c|}{ICUMM}        & 29.54			& 53.59			& 34.45      & 18.81         & 0.1427        \\ \hline
		\multicolumn{2}{|c|}{Mean-Variance Model}		& 64.14		& 89.87		& 31.79		&	19.06		& 0.1737	\\ \hline
		\multicolumn{7}{|c|}{Patient 6}                             \\ \hline
		\multicolumn{2}{|l|}{} &1-std \% & 2-std \% & RMSE & MPE & CORR \\ \hline
		\multirow{2}{*}{MSG Model}	& optimization  & 59.04			& 87.45			 & 43.16      & 25.22         & 0.2859        \\ \cline{2-7}
		& MCMC		& 63.84		& 91.14		& 44.78		& 27.12		& 0.1859		\\ \hline
		\multicolumn{2}{|c|}{ICUMM}        & 18.45			& 38.01			& 44.77      & 26.18         & 0.1374        \\ \hline
		\multicolumn{2}{|c|}{Mean-Variance Model}		& 63.84		& 91.14		& 46.66		& 28.04		& 0.1231 \\ \hline
	\end{tabular}
	\caption{Comparison of the forecasting results obtained with the MSG, ICUMM and mean-variance models. The percentages of 1- and 2-stdev bands that capture the true BG measurements with the MSG model is substantially higher than the ICUMM whereas they are smaller than mean-variance model. On the other hand, RMSE and MPE values are closer when comparing all three models, yet the MSG model still provides smaller values for these measures, as well. In addition, the forecasted BG levels by MSG model gives the highest correlation with the BG measurements for all three patients.}
	\label{tab4}
\end{table}

The comparison results are shown in Table \ref{tab4}.
These results indicate the efficiency of the MSG model in forecasting
BG measurements and assessing the uncertainty in the forecasted values.

\emph{First}, the point estimators in
the MSG case exhibit comparable, or improved, accuracy in comparison
to the ICUMM and mean-variance model, i.e.,  RMSE and MPE are smaller for the MSG
model obtained either via optimization or MCMC. In addition, correlation
values obtained with the MSG model are significantly larger than those with
the ICUMM and mean-variance
model. These results show that with a relatively simple model,
we are able to reach the same, or better, accuracy in forecasting BG
behavior than a more physiologically based high-fidelity model, with a
larger number of unknown model parameters.

\emph{Second}, the confidence
bands that we use to quantify possible high and low values of BG level
could provide better results than ICUMM. Compared to ICUMM, the
improved accuracy of the MSG model in terms of uncertainty forecasting may be
related, in part, to the fact that the model we use is inherently stochastic, and fits
the stochastic  fluctuations to data; in contrast, ICUMM provides uncertainty bands
only through the ensemble of solutions which are a product of the algorithm used
to fit the data, and not inherent to the model itself. On the other hand, these
2-stdev bands contain only slightly larger number of BG measurements with
mean-variance model compared to MSG model. Even though mean-variance model
provides better results with respect to this evaluation metric, it gives larger RMSE
and MPE and much smaller correlation between the BG measurements. Besides, the
mean-variance model does not provide any physiological understanding of the system.

In summary, a comprehensive comparison of MSG model with a more complex
physiology-based mechanistic model and a simpler data-driven model
suggests that the MSG model, a mechanistic model with small number of
parameters, works at least as good as, if not better than, these models in
representing BG behavior and forecasting future BG levels in the ICU setting.

\section{Discussion and Conclusion}\label{conc}

\textbf{Summary of the modeling framework:} In this paper, we introduce a new mathematical model of the glucose regulatory system in humans. The model was created with five goals in mind: \emph{(i)} the model should be robustly identifiable/estimable and verifiable with real world human data \cite{sherman2016real}---data collected for health management---such that the model could potentially be useful for personalized parameter estimation and state forecasting \cite{sherman2016real};  \emph{(ii)} the model should be interpretable in the sense that patient specific parameters may be used to explain, and quantify basic physiological mechanisms; \emph{(iii)} the model should be physiologically simple, even if the model is functionally complex, to minimize parameter identifiability problems present in many existing physiological models;  \emph{(iv)} a model framework generalizable and adaptable to several contexts including T2DM and ICU; and \emph{(v)} a model that was amenable to a model-based control environment. With these goals driving the model development, our model follows different approach, as explained in Section \ref{model_construct} compared to many other glucose-insulin modeling efforts. The most important departure of our model compared with others is the inclusion of insulin as a lumped parameter affecting the glucose state rather than as an independent state or state(s). We formulated the model this way because, in clinical settings, insulin is rarely measured, and therefore difficult to estimate.

\textbf{Model development constrained by real world data:} Restricting model development to the constraints imposed by readily available real world data is a severe, but important, restriction. To be directly useful in applications, models must be estimable using data that are collected within the context of the given application, and these data are almost always much more sparse than ideal laboratory experiments. To circumnavigate these problems, we are forced to use data collected to manage health and the models that can be applied with these data will likely be different than models built to be estimable with, e.g., laboratory data. Therefore, to help facilitate the circular process of allowing our knowledge of systems physiology to inform and impact how clinicians manage the health of people, we need a bridge between these worlds, and the bridge proposed here is through inference with data based on simple yet interpretable models.

\textbf{Including insulin in a mechanistic modeling framework:} There are, at a high level, two pathways for estimating and modeling blood glucose behavior. First, one can include dynamic equations for glucose and insulin, along with other related processes. Second, one can include a dynamic equation for glucose that includes a parameterized function capturing the impact of insulin on glucose that is not dynamic. The second option excludes a model equation for insulin and instead has a parameterized function that represents the impact of glucose on insulin that is hypothesized to exist. The most common tactic is the first approach. However, as we emphasized before, insulin measurements rarely exist in any practical setting and insulin levels are not interpretable or meaningful for patients or for most clinicians. This implies that modeling glucose together with insulin leaves one of two primary variables free to vary, causing the system to be ill-posed and not identifiable. In contrast, the second option does not have these pathologies, will be uniquely estimable under most circumstances, and will provide a more stable forecast.

\textbf{Blood glucose variability and its effect on uncertainty quantification:} Both in T2DM and ICU settings, we obtain the forecasted mean and 2-stdev bands around this mean as our model output. The characteristics of BG behavior, in particular BG variability, affect the width of the 2-stdev bands in these two settings. It is natural to expect to have wider bands in the ICU setting because of the highly non-stationary BG behavior of ICU patients. However, there are other factors affecting the width of these bands. First, BG measurements are relatively dense in the ICU, compared to T2DM. The second factor is the difference between the experiments that were designed according to the specific needs of each setting, which is depicted in Fig~\ref{fig1} and \ref{fig2}. Using the latest 24-hour data for model estimation and forecasting in the ICU setting could help reduce uncertainty in the forecasts and may result in occasional more narrow 2-stdev bands. In addition, the GM is performed by clinicians in ICU and by patients in the T2DM setting. A clinician's management is likely to reduce the variability in BG behavior, which can be reflected in the width of the estimated 2-stdev bands.

\begin{figure}[ht]
	\centering
	\includegraphics[scale=0.4]{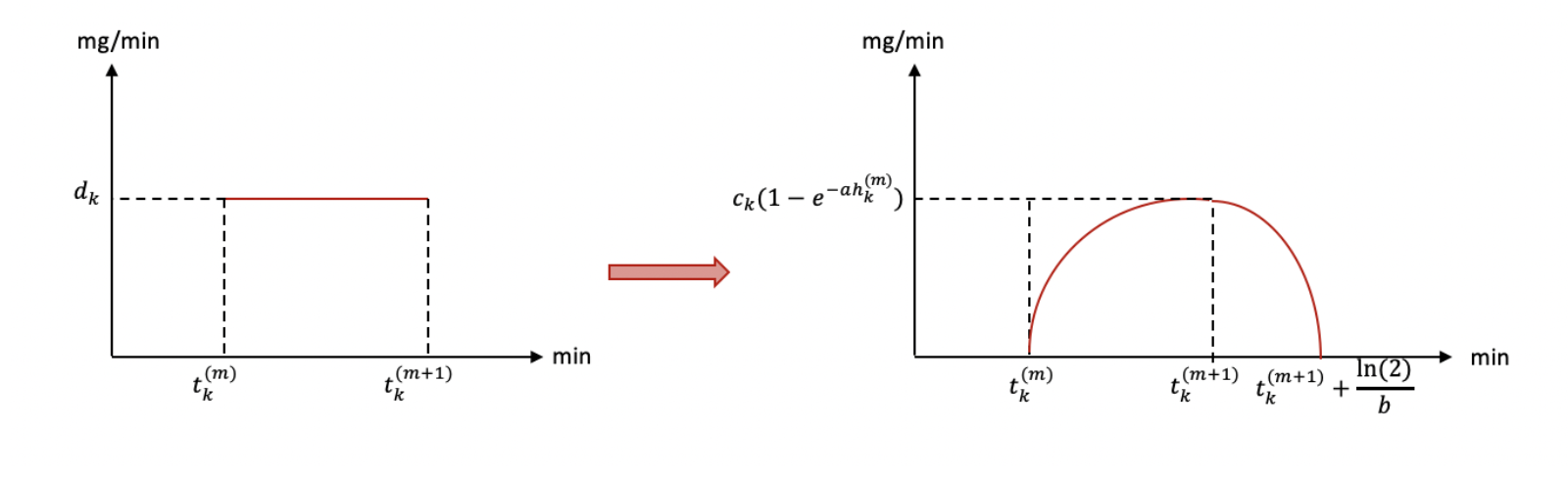}
	\caption{Smoothing piecewise constant nutrition function that is used for ICU patients}
	\label{fig8}
\end{figure}

\textbf{Impact of the nutrition function choice in the ICU context:} We also considered a different form for the nutrition function in the ICU setting to test robustness of our modeling to the simplistic piecewise constant meal function that we adopt in this case. Because ICU patients are tube-fed with nutrition quantities that are considerably less than a healthy individual would ingest, per unit time,  it is reasonable to consider the use of a piecewise constant function. Nonetheless we investigated if modifying the piecewise constant function as shown in Fig~\ref{fig8} could improve the parameter estimation and/or forecasting results. Using this function introduces two more parameters to be estimated in the ICU setting, increasing the flexibility and the complexity at the same time. Hence the parameters to be estimated are $G_b,\gamma,\sigma,\beta,a,b$. Using this function modeling the nutrition delivery did not improve the forecast accuracy which lead us to use the simpler version of the model.

The function shown on the right hand side of Fig~\ref{fig8} is formulated as follows:

\begin{align}
	\begin{aligned}
		m(t) = \sum_{k=1}^{K_m}c_k&((1-e^{-a(t-t_k^{(m)})})\1_{[t_k^{(m)},t_{k+1}^{(m)})}(t)\\
		&+(1-e^{-ah_k^{(m)}})(2-e^{b(t-t_k^{(m+1)})})\1_{[t_k^{(m+1)},t_{k+1}^{(m)}+\frac{ln(2)}{b}]}(t)),\label{exp_nut_icu}
	\end{aligned}
\end{align}
where $h_k^{(m)}:=t_k^{(m+1)}-t_k^{(m)}$ and
$$c_k := \frac{d_kh_k^{(m)}}{h_k^{(m)}+(e^{-ah_k^{(m)}}/a)+(1-e^{-ah_k^{(m)}})((2ln(2)-1)/b)},$$ is the normalizing constant for $k=1,2,...,K_m$.

\textbf{Outlook:}\label{outlook} The model we have developed has demonstrable predictive capability and discriminates between datasets in a patient-specific manner. Yet it has some limitations, which give space for future development, and also suggests some natural next-step applications. We outline a number of possible future directions.

\emph{Glycemic control:} Given the MSG model construction, an obvious next step is to formulate the work on the control problem where we determine estimates of
the input ranges of nutrition and insulin, necessary to keep the output, here BG, in a desired target range. 

\emph{Phenotyping:} Because the parameters of the MSG model are interpretable and track physiology reasonable well, we could potentially use the model parameter estimates for phenotyping studies, \cite{albers2014dynamical,albers2012population,albers2018mechanistic}. Meaning, we could estimate parameter for individuals in a given health state, establishing an inferred phenotype for the patient, and then relate this phenotype to other external health features or cluster the patient phenotypes in an effort to find structure among the inferred physiology. We have deemed efforts such as this high-fidelity phenotyping \cite{hripcsak2017high} and believe that this model has the potential to be used to these ends. For example, one of the clinical criteria to diagnose diabetes is fasting blood glucose levels higher than 126 mg/dl, where fasting is defined as no caloric intake at least eight hours. We intend to understand the relationship between the fasting blood glucose level and our model parameter, $G_b$, the basal glucose value. The existence of such relationship could show us if $G_b$ is useful to identify diabetes.

\emph{Further model generalization to include other glucose-data driven situations:} We have not investigated how the MSG model might work given oral glucose tolerance test (OGTT) data. The OGTT is one of the standard settings for glucose-insulin model development and potential use; we know of only one model that currently generalized to both OGTT and clinical data \cite{ha2015mathematical} and we would like to add the MSG model to this list. 

\emph{T1DM:} We have an initial version of the MSG model that could be used within T1DM setting. It would be interest to test this version on T1DM data. Since the time-scales of health progression here are more similar to those of T2DM than the ICU setting, giving hope that the method might have similar predictive capability in this setting. While MSG model may not outperform current T1DM ODE models for short-term glucose forecasting or control, it may useful for performing larger-scale, longitudinal patient phenotyping. Because of not having access to such a T1DM dataset, we have not been able to work with this version in this paper. We plan to pursue a number of the research directions outlined here in the immediate future.

\printbibliography

\end{document}

\begin{flushleft}
	{\Large
		\textbf{Supplementary Material}
	}
\end{flushleft}

\subsection*{Optimization}\label{optimization}

In solving the parameter estimation problem with the optimization approach, our goal here is to determine parameter values, $\theta$, which maximize the posterior distribution, $\mathbb{P}(\theta|y)$ and is called to be the {\em maximum a posteriori (MAP) estimator}. Using the prior distribution as specified above, the parameter estimation problem becomes
\begin{equation}
	\theta^* = \argmax_{\theta}\mathbb{P}(\theta|y) = \argmax_{\theta\in\Theta}\mathbb{P}(y|\theta) = \argmin_{\theta\in\Theta}-\log(\mathbb{P}(\theta|y)).\label{opt_param_est_1}
\end{equation}
Remember that for the parameter estimation problem with the Minimal Stochastic Glucose model, we have
\begin{equation}
	-\log(\mathbb{P}(\theta|y)) = \frac{K_m}{2}\log(2\pi)+\frac{1}{2}\log(\det(S(\theta)))+\frac{1}{2}(y-Lm(\theta))^TS(\theta)^{-1}(y-Lm(\theta)).\label{Log_Likelihood}
\end{equation}
Then, substituting \eqref{Log_Likelihood} into \eqref{opt_param_est_1}, the problem will take the form

\begin{equation}
	\theta^* = \argmin_{\theta\in\Theta}||S(\theta)^{-1/2}(y-Lm(\theta))||^2+\log(\det(S(\theta))).\label{opt_param_est_2}
\end{equation}
Hence, placing uniform prior distribution turns the problem of finding the MAP estimator into a constrained optimization problem. To solve this problem, we use built-in {\ttfamily MATLAB} functions, such as {\ttfamily fmincon} and {\ttfamily multistart}. {\ttfamily fmincon} is a gradient-based minimization algorithm for nonlinear functions. {\ttfamily multistart} starts the optimization procedure from the indicated number of starting points that are picked uniformly over the region defined by the constraints. It uses {\ttfamily fmincon} and other similar algorithms to perform each optimization process independently and provides the one that achieves the minimum value among the results of all separate runs. With this approach, we have the opportunity to compare different optimization procedures that start from different initial points. This provides some intuitive understanding of the solution surface and hence the estimated optimal parameters.

Once an optimal point has been found, we may also employ the
Laplace approximation \cite{mackay2003information,owen2013monte}
to obtain a Gaussian approximation to the posterior distribution.
The Laplace approximation is a reasonable approximation in many data
rich scenarios in which all parameters are identifiable from the
data, because of the Bernstein Von Mises Theorem \cite{van2000asymptotic},
which asserts that the posterior distribution will then be approximately
Gaussian, centered near the truth and with variance which shrinks to
zero as more and more data is acquired. However data is not always
abundant, and not all parameters are identifiable even if it is; in this
setting sampling the posterior distribution is desirable, and for this
purpose, we use Markov Chain Monte Carlo (MCMC) techniques.

\subsection*{Markov Chain Monte Carlo}\label{mcmc}

MCMC methods are a flexible set of techniques which may be used to sample from a target distribution, which is not necessarily analytically tractable, \cite{liu2008monte,robert2013monte}. For example, the distribution $\mathbb{P}(\theta|y)$ is the conditional distribution of the random model parameters, $\theta$ given the data, $y$. Even though we can explicitly formulate it using the Bayes' Theorem, it is not always an easy task to extract useful quantities, such as posterior mean and variance, from that formula. In such cases, MCMC techniques are used to generate random samples from this target distribution and this random sample is used to obtain the desired information, which could be anything such as the mean, mode, covariance matrix, or higher moments of the parameters. Moreover, this technique is also very helpful to obtain uncertainty quantification (UQ) results for the estimated parameters.

In order to obtain more extensive knowledge than MAP estimator can provide about the posterior distribution of parameters given the data, $\theta|y$, we use MCMC methods as a natural choice to sample from that distribution. Among different possible algorithms (see \cite{gelman2013bayesian}), we use the standard random walk Metropolis-Hastings algorithm. In order to make sure the resulting sample is indeed a good representer of the posterior distribution, we perform some diagnostics such as checking if chains for each parameter converged and if they are uncorrelated. Then, after removing the burn-in period, we compute the mean and the covariance matrix from the remaining part of the sample. We use the mean as a point estimator for simulation and forecasting, and the covariance matrix provides valuable information to quantify uncertainty for the estimated parameters.

In general, it is hard to obtain efficient results with MCMC methods even when sampling from the joint distribution of four or five parameters, due to the issues such as parameter identification. Moreover, obtaining accurate results with this approach requires careful choice of starting point and tuning some other parameters. In general the performance of the algorithm will depend on the initial point. We tested the use of both random starting points and MAP estimators as starting point. The former enables us to detect when several modes are present in the posterior distribution; the latter helps to focus sampling near to the most likely parameter estimate and to quantify uncertainty in it. However, it is also important to note that using MAP estimator as a starting point is not helpful in all cases. More precisely, if the MAP estimator is not a global minimum but a local minimum, then the chain could get stuck around this point. Therefore, it requires careful analysis, comparison and synthesis of the results obtained with these different approaches.

\subsection*{Insight About the MSG Model's Ability to Capture ICU Dynamics}\label{insight_icu}

\begin{figure}[!ht]
	\begin{subfigure}[b]{0.33\textwidth}
		\centering
		\includegraphics[width=\linewidth]{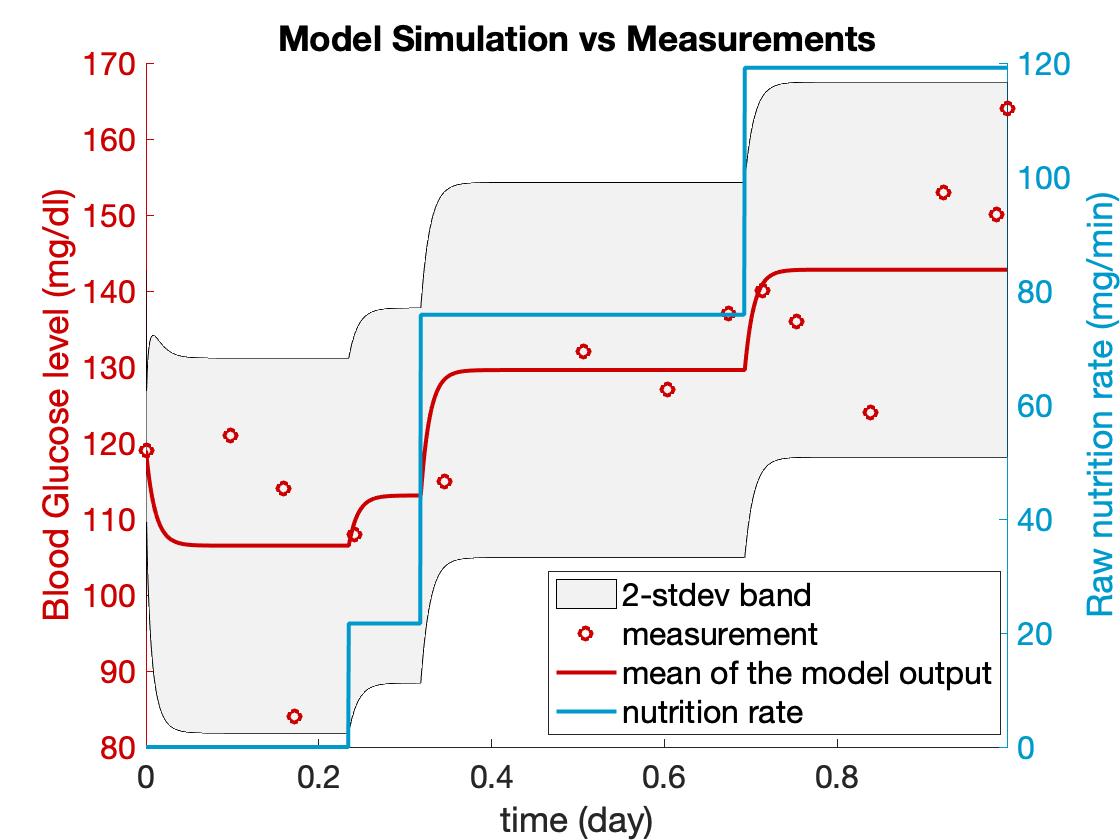}
		\captionsetup{justification=centering}
		\caption{$G_b$=106.49,\\ $\gamma$=0.07, $\sigma$=12.34}
		\label{fig9A}
	\end{subfigure}%
	\begin{subfigure}[b]{0.33\textwidth}
		\centering
		\includegraphics[width=\linewidth]{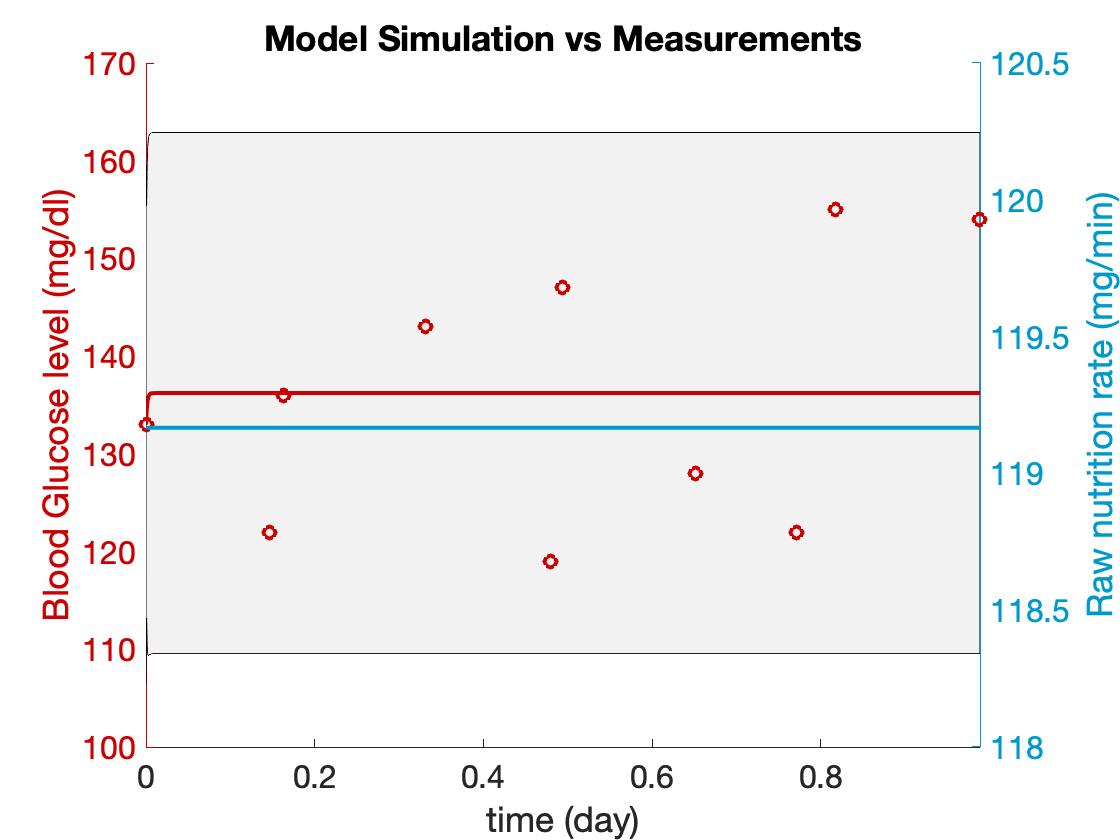}
		\captionsetup{justification=centering}
		\caption{$G_b$=131.40,\\ $\gamma$=0.49, $\sigma$=13.33}
		\label{fig9B}
	\end{subfigure}%
	\begin{subfigure}[b]{0.33\textwidth}
		\centering
		\includegraphics[width=\linewidth]{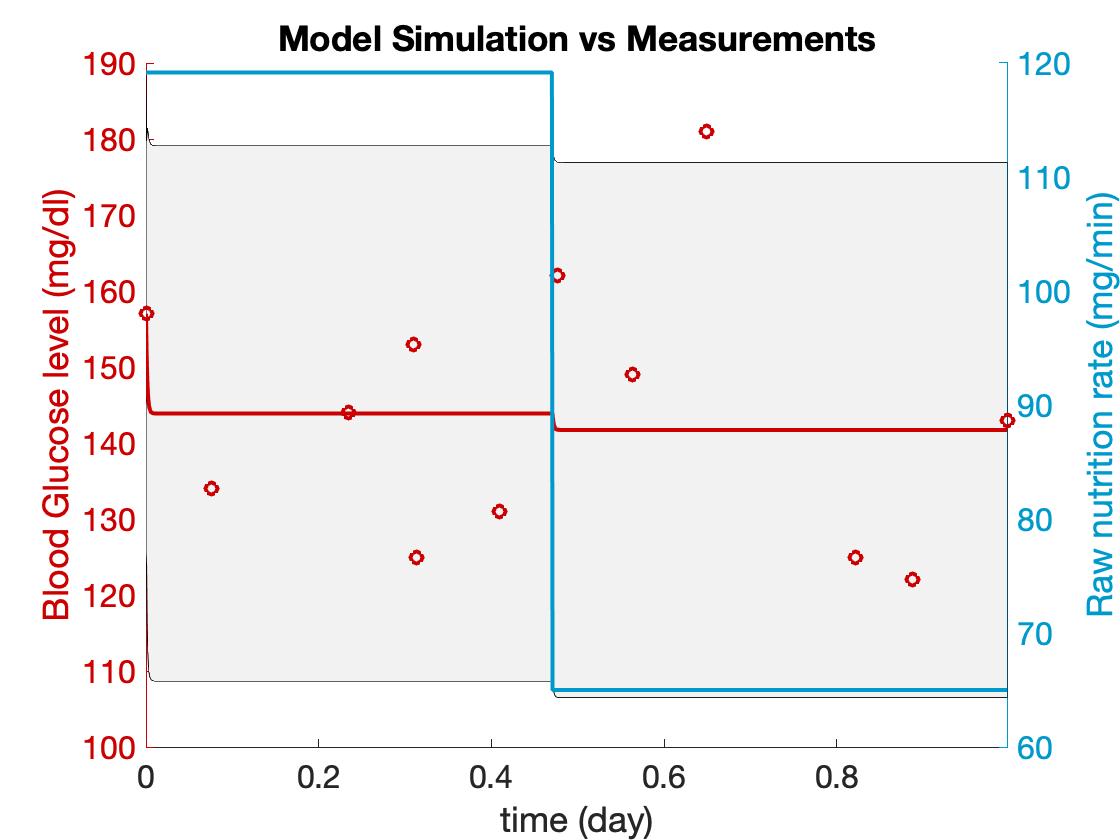}
		\captionsetup{justification=centering}
		\caption{$G_b$=139.12,\\ $\gamma$=0.50, $\sigma$=17.61}
		\label{fig9C}
	\end{subfigure}%
	\caption{BG simulations are shown with respective to the estimated parameters over respective training time window. In each plot, the light blue curve is the glucose rate in the nutrition delivered to the patient (right y-axis), the red circles show the true BG measurements (left y-axis), the red curve is the mean of the model output (left y-axis), and the gray area is the 2-stdev band around the mean of the model output (left y-axis). These figures show two main cases that could arise as a result of parameter estimation in the ICU setting: \textbf{(a)}: the input (nutrition rate) is reflected in the output (BG measurements), \textbf{(b)} and \textbf{(c)}: the input is not reflected in the output, which makes it impossible to estimate the decay rate $\gamma$ accurately.}
	\label{fig9}
\end{figure}

We explain here how the model captures the BG dynamics in the ICU setting. Consider Fig~\ref{fig9}, which demonstrates both the model's relative robustness and its capability of capturing the dynamics and various complexities encountered in different conditions in the ICU setting. These figures show simulated BG values for patient 4 over different training time windows for which the parameters are estimated with the optimization approach. Here the the red curves represent the mean of the BG dynamics that are assumed to be oscillatory, the amplitude of oscillations are expected to lie in the gray region as it is the 2-stdev band around the mean. The red circles show the BG measurements the light blue curve shows the tube-nutrition input rate. For simplicity we consider patient 4 who did not need any exogenous insulin, so the tube-feed nutrition is the only driver of the BG level. Each subfigure of Fig~\ref{fig9} shows a different training time window that is representative of different circumstances relative to our ability to estimate $G_b$, $\gamma$, and $\sigma$.

Fig~\ref{fig9A} shows a situation where the BG measurements (the output of the system) reflect the nutrition rate (the input to the system) quite well. As the nutrition rate, the only driver of the system included in the model, the BG measurements increases, which is reflected in the mean of the model output.

In contrast, Fig~\ref{fig9B} and \ref{fig9C} demonstrate a situation where the BG measurements do not reflect the nutrition rate over the shown time window. In Fig~\ref{fig9B}, we see that there is no change in the nutrition rate over the whole time window, which is clearly reflected by the mean of the model output. However, having no change in the nutrition rate means that there is no opportunity to ``learn" the glycemic decay parameter, $\gamma$. Therefore, in the absence of any information, optimization algorithm provides the possible largest value to reflect non-changing nutrition rate in the mean of the model output, which also around the mean of all the BG measurements in the respective training time window.  

On the other hand, the situation in Fig~\ref{fig9C} occurs when the changes in the BG measurement are uncorrelated with the changes in the nutrition rate, potentially due to changes in health states or other interventions, e.g., other hormone drips. Hence it is impossible for the model parameters to accurately reflect the physiology as they are accounting for dynamical glucose features that they were not designed to accommodate. Observe from Fig~\ref{fig9C} that, with a $\gamma$ value as in Fig~\ref{fig9A}, there should be a substantial decrease in the mean of the model output due to the decrease in the nutrition rate. Since this is not the case as seen by the BG measurements, the optimization algorithm provides a reasonable mean model output (characterized by $G_b$) and set the decay rate as high as possible (here the upper bound for this parameter) so that it could keep the mean model output constant over the whole interval accommodating for the two widely different nutrition rate regimes.

These issues do not mean the model cannot represent and forecast the glycemic dynamics, it still is usually able to represent glycemic dynamics, but \emph{some of the parameters} might lose their intended meaning. For example, in the two respective examples, despite parameter estimate issues, in both of these cases the estimated $G_b$, and $\sigma$ values are enough to capture the mean and variance of the BG measurements accurately. These examples are not the only cases where we observe parameter estimates that are not physiologically valid while at the same time the glucose forecast and modeling itself remains accurate. The other examples are all variations of the same theme; we either do not have the available data to estimate a parameter accurately, or the data are behaving in a more complex manner, and in both cases, the parameters make up for these data-driven and model-driven short-comings by deviating from their normal roles to render a robust glucose forecast. It is likely that problems such as these will not be eliminated by using more complex datasets and more complex models, because full representation of the relevant processes is out of reach in such non-stationary ICU settings.

\subsection*{Parameter Identifiability Issues in the ICU Setting}

\begin{figure}[!ht]
	\begin{subfigure}[b]{\textwidth}
		\centering
		\includegraphics[width=.7\linewidth]{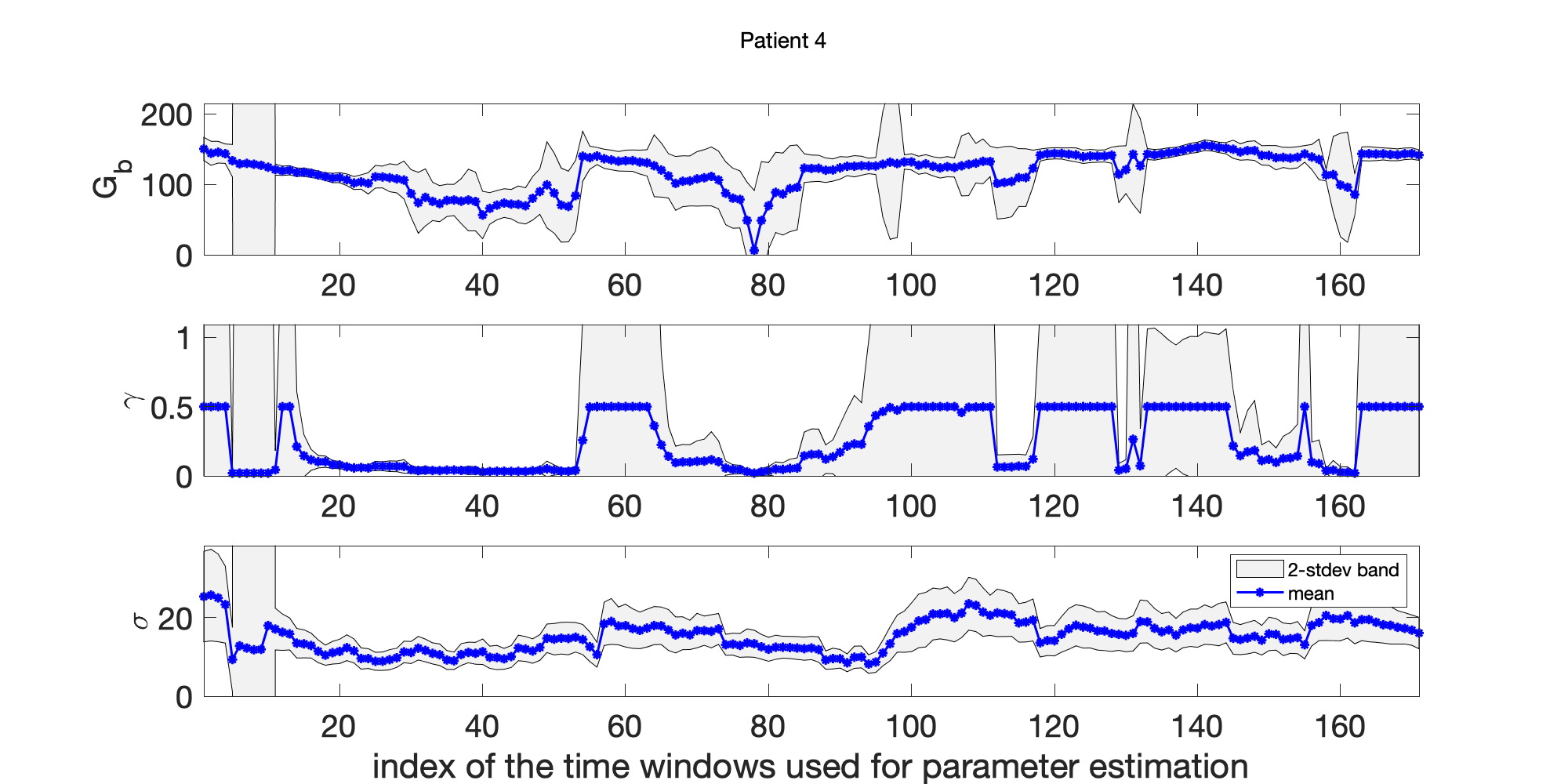}
		\captionsetup{justification=centering}
		\caption{Patient 4}
		\label{fig10A}
	\end{subfigure}

	\begin{subfigure}[b]{\textwidth}
		\centering
		\includegraphics[width=.7\linewidth]{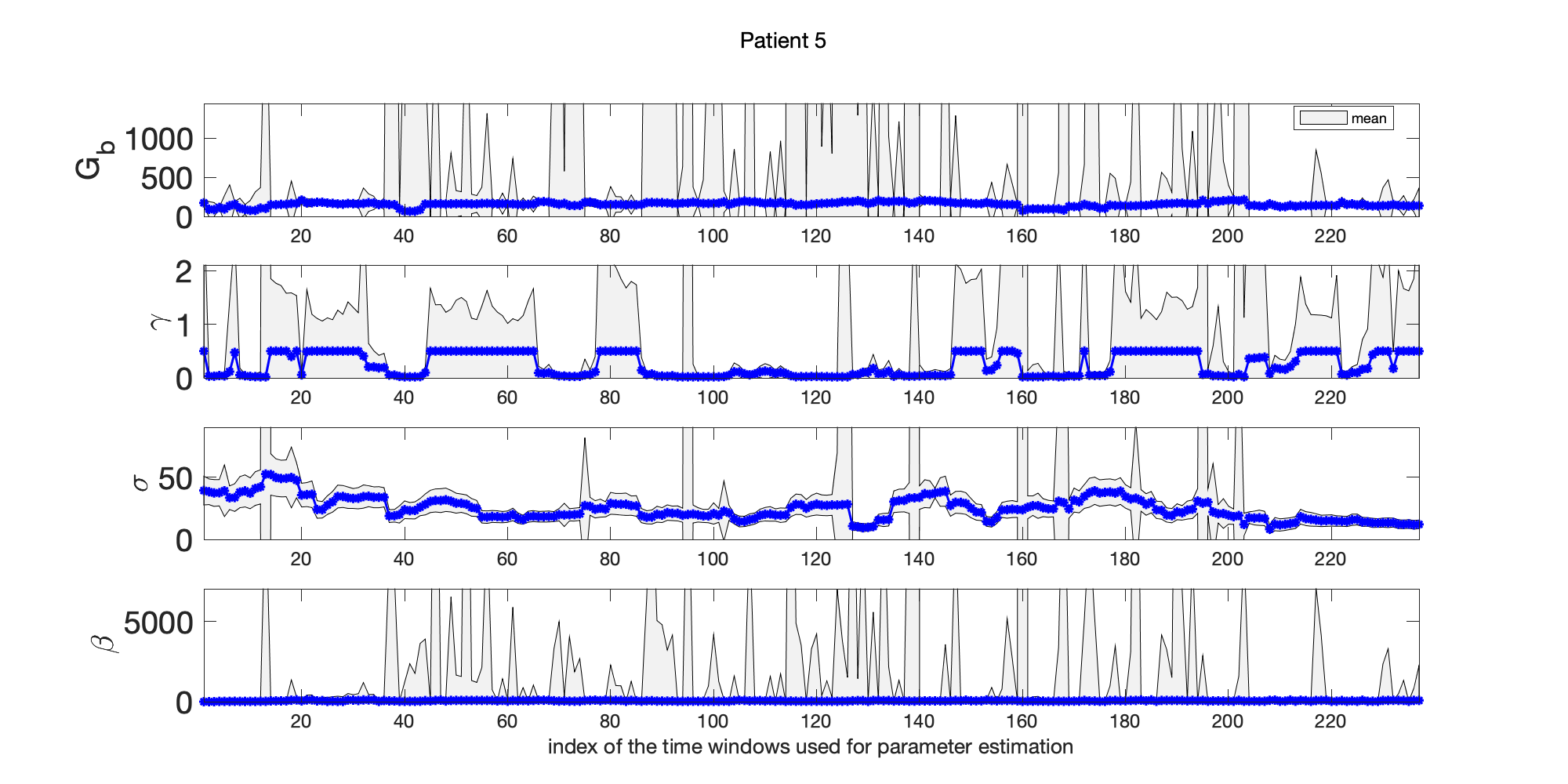}
		\caption{Patient 5}
		\label{fig10B}
	\end{subfigure}
	
	\begin{subfigure}[b]{\textwidth}
		\centering
		\includegraphics[width=.7\linewidth]{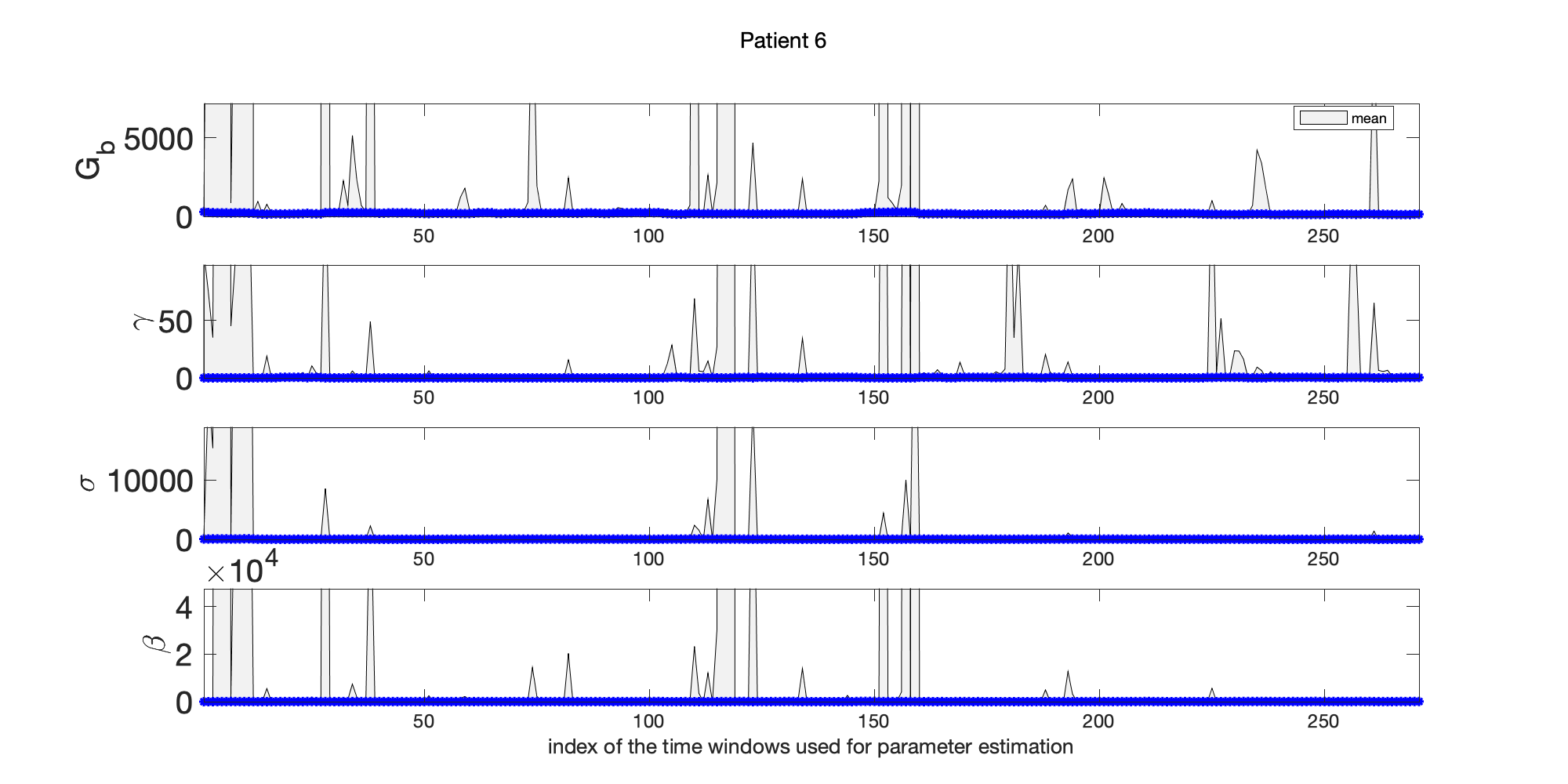}		
		\caption{Patient 6}
		\label{fig10C}
	\end{subfigure}

	\caption{The figure shows the UQ bands when we use the optimization approach for parameter estimation in the ICU setting. Because of the parameter identifiability issues, the Hessian matrix is-ill conditioned. When we compute its inverse to obtain the variance in the estimated model parameters, we obtain unreasonably wide UQ bands.}
	\label{fig10}
\end{figure}

We encountered some parameter identifiability issues while solving the optimization problem stated in \eqref{opt_param_est_2} in the ICU setting, which caused the solution surface, the function we minimize ($\mathbb{P}(\theta|y)$), to be flat in some parameter directions. Because of this reason, the Hessian matrix becomes ill-conditioned. However, the Laplace approximation requires computing the inverse of the Hessian matrix, which gives the variance of the estimated unknown model parameters on the diagonal. When we compute the inverse of an ill-conditioned Hessian matrix, the resulting matrix has very large entries. Hence the variance of the estimated model parameters is also unreasonably large, resulting in non-useful UQ results, which can be seen in Fig~\ref{fig10}.

\subsection*{The Effect of Blood Glucose Measurement Pattern to Parameter Estimation in the T2DM Setting}

\begin{figure}[!ht]
	\begin{subfigure}[b]{0.33\textwidth}
		\includegraphics[width=\linewidth]{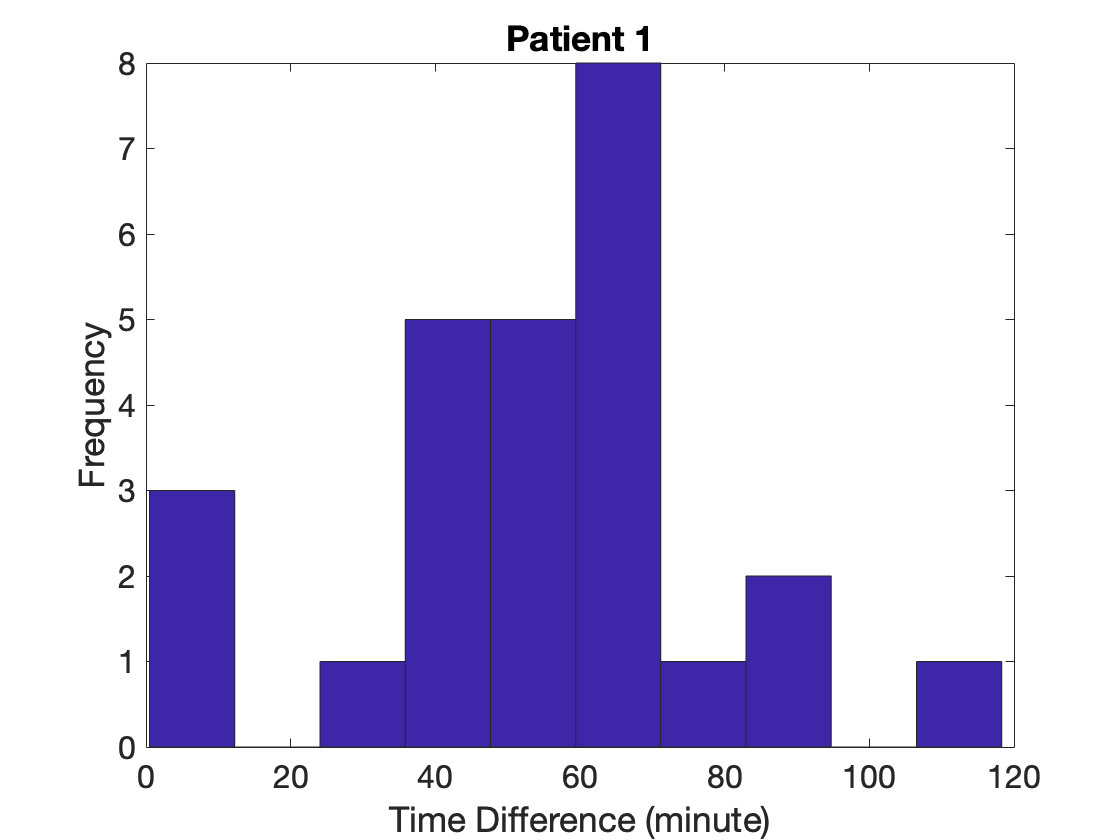}
		\caption{Patient 1}
		\label{fig11A}
	\end{subfigure}%
	\begin{subfigure}[b]{0.33\textwidth}
		\includegraphics[width=\linewidth]{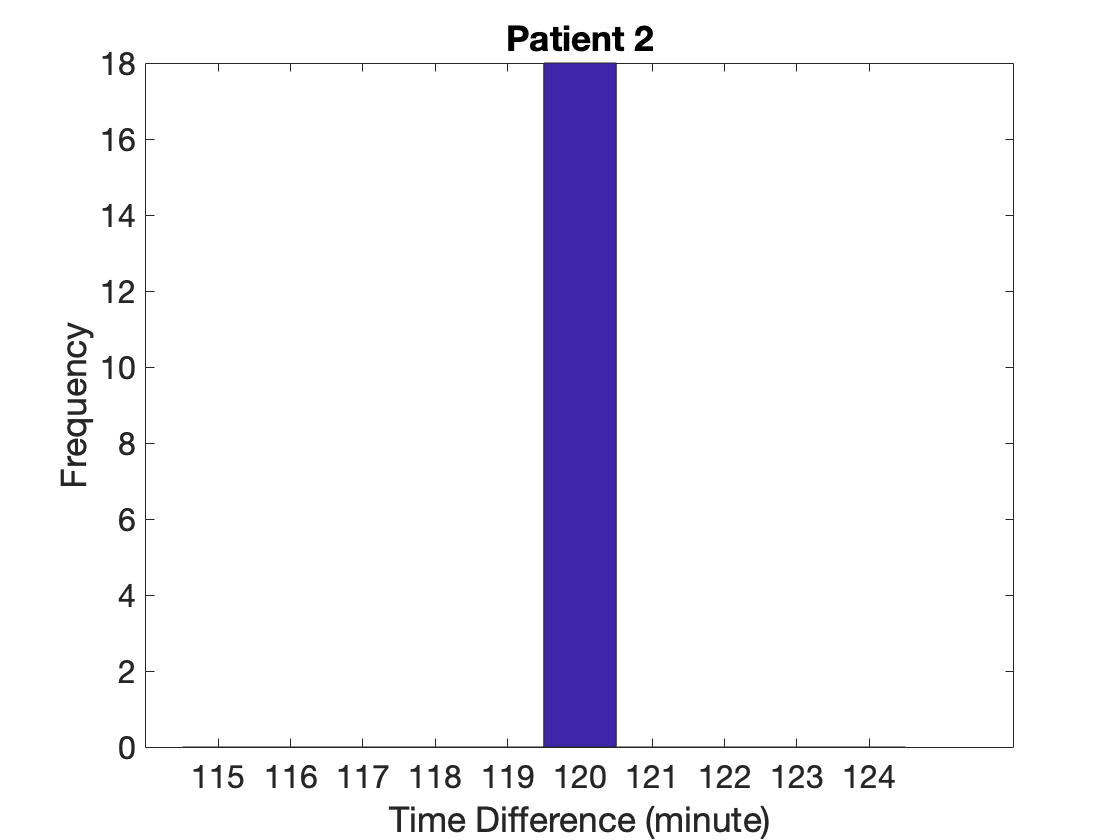}
		\caption{Patient 2}
		\label{fig11B}
	\end{subfigure}%
	\begin{subfigure}[b]{0.33\textwidth}
		\includegraphics[width=\linewidth]{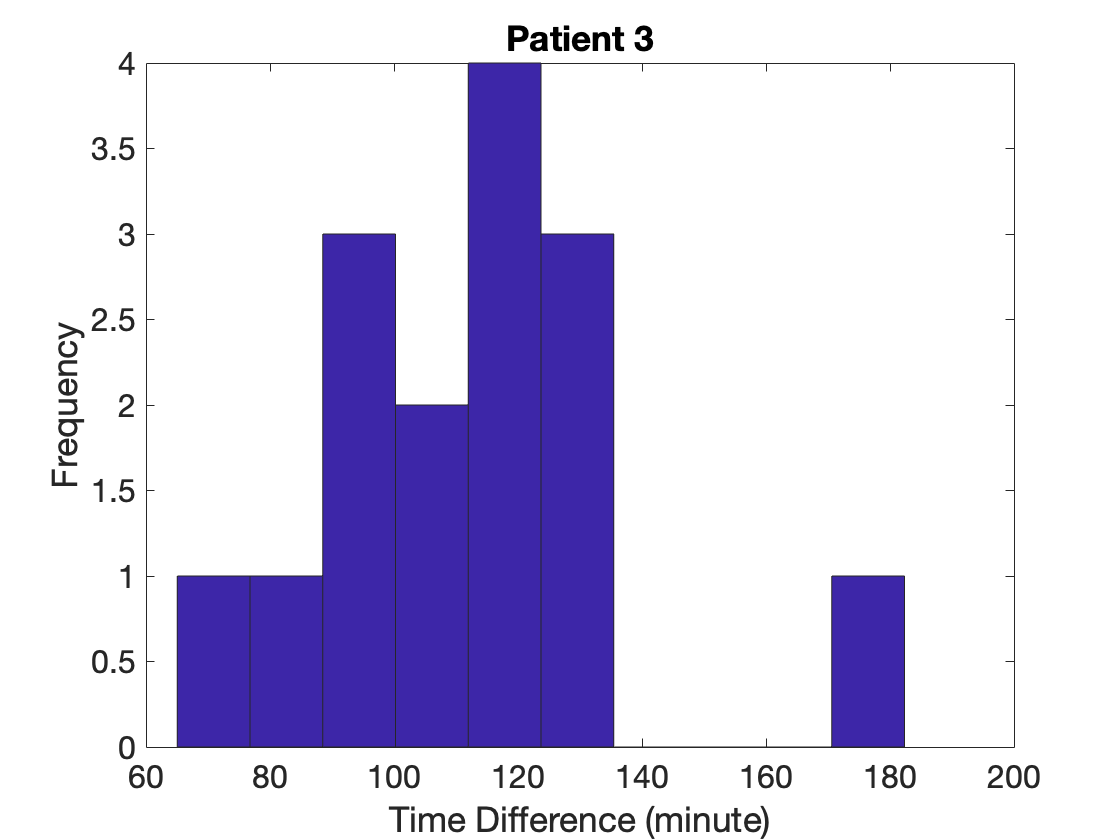}
		\caption{Patient 3}
		\label{fig11C}
	\end{subfigure}%
	
	\caption{Each histogram shows the elapsed time after meal time until the first BG measurement for each meal in the training time window. Lack of variability of patient 2 causes suboptimal parameter estimation and forecasting results for this patient.}
	\label{fig11}
\end{figure}

We observed that the BG measurement pattern of T2DM patients significantly affects the values of estimated model parameters based on this patient-collected data. For each patient, we computed the time difference between the meal time and the first BG measurement after that meal time for each meal recorded during the training time window. We show the histogram of these time differences for each patient in Fig~\ref{fig11}. We see that patients 1 and 3 measured their BG levels at varying times after meal consumption. However, patient 2 measured precisely after two hours for each meal.

\FloatBarrier

\printbibliography